\documentclass[11pt]{article}

\usepackage{natbib}
\usepackage{amsmath,amsfonts,latexsym,amssymb,graphicx}
\DeclareGraphicsExtensions{.pdf,.gif,.jpg,.png}

\newcommand{\argmax}{\ensuremath{\operatornamewithlimits{argmax}}}

\newcommand{\fbold}    {\mbox{\bf f}}

\newcommand{\ubold}    {\mbox{\bf u}}

\newcommand{\xbold}    {\mbox{\bf x}}
\newcommand{\Xbold}    {\mbox{\bf X}}
\newcommand{\ybold}    {\mbox{\bf y}}
\newcommand{\Ybold}    {\mbox{\bf Y}}
\newcommand{\zbold}    {\mbox{\bf z}}

\newcommand{\epsilonbold} {\mbox{\boldmath${\epsilon}$}}

\usepackage{times}
\usepackage{soul}
\usepackage{epic,eepic}
\usepackage{multicol}
\usepackage{xfrac}

\newcommand{\subheading}[1]{\medskip\noindent{\em{#1}}}

\newcommand{\var}{\mbox{Var}}
\newcommand{\R}{{\mathbb R}}
\newcommand{\Z}{{\mathbb Z}}
\newcommand{\dt}{\mbox{\small$\Delta$}{t}}

\begin{document}

\title{Control Theory and Experimental Design in Diffusion Processes}


\author{Giles Hooker\thanks{Department of Biological Statistics and
    Computational Biology, Cornell University} \and Kevin
  K.~Lin\thanks{Department of Mathematics and Program in Applied
    Mathematics, University of Arizona} \and Bruce
  Rogers\thanks{Department of Mathematics, Southwest Illinois College}}

\date{}

\maketitle


\begin{abstract}
This paper considers the problem of designing time-dependent, real-time
control policies for controllable nonlinear diffusion processes, with
the goal of obtaining maximally-informative observations about
parameters of interest.  More precisely, we maximize the expected Fisher
information for the parameter obtained over the duration of the
experiment, conditional on observations made up to that time.  We
propose to accomplish this with a two-step strategy: when the full state
vector of the diffusion process is observable continuously, we formulate
this as an optimal control problem and apply numerical techniques from
stochastic optimal control to solve it.  When observations are
incomplete, infrequent, or noisy, we propose using standard filtering
techniques to first estimate the state of the system, then apply the
optimal control policy using the posterior expectation of the state.  We
assess the effectiveness of these methods in 3 situations: a
paradigmatic bistable model from statistical physics, a model of action
potential generation in neurons, and a model of a simple ecological
system.
\end{abstract}


%
%
%

\section{Introduction} \label{sec:intro}

This paper considers a novel problem in experimental design, that of adaptively choosing
inputs in stochastic systems to yield more precise estimates of system parameters. We suppose an
experimenter is studying a physical or biological system that can be
modeled by a diffusion process of the general form
\begin{equation}
  d\xbold_t = \fbold(\xbold_t,\theta,u_t) dt + \Sigma(\xbold_t)^{1/2}
  dW_t~.
  \label{eq:sde}
\end{equation}
Here, $\xbold_t$ is the state vector the system at time $t$, which
evolves according to a (deterministic) vector-valued drift function
$\fbold(\xbold_t,\theta,u_t)$ that depends on an unknown parameter
$\theta$,
as well as a time-varying input $u_t$ under the direct control of the
experimenter.
In addition to the drift $\fbold$, the system is also influenced by
noise, modeled as a multivariate Brownian motion $dW_t$ with a
state-dependent covariance matrix $\Sigma(\xbold,t)$~.
The functional forms of $\fbold$ and $\Sigma$ are assumed known -- usually from first principles -- and
experiments are to be conducted to estimate the value of $\theta$. We consider the problem: How
can the experimenter design a protocol for setting the control signal
$u_t$ dynamically, in real time, using all the information available up to time $t$ to optimize the precision of our estimate of $\theta$? To accomplish this, we use the tools of dynamic programming in order to account not only for the information gained from the future evolution of the system, but also the future control policies that will be applied to it.



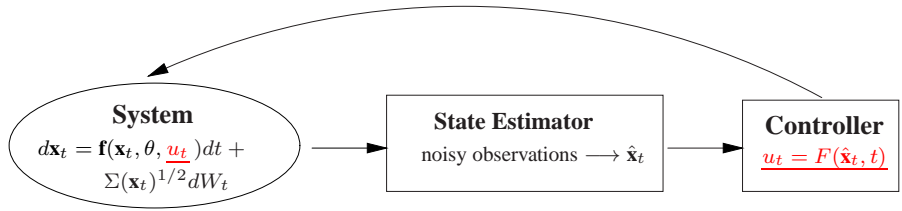
\begin{figure}
\setlength{\unitlength}{0.00083333in}
\begingroup\makeatletter\ifx\SetFigFont\undefined%
\gdef\SetFigFont#1#2#3#4#5{%
  \reset@font\fontsize{#1}{#2pt}%
  \fontfamily{#3}\fontseries{#4}\fontshape{#5}%
  \selectfont}%
\fi\endgroup%
{\renewcommand{\dashlinestretch}{30}
\begin{picture}(5570,1158)(0,-10)
\path(4118,381)(4583,381)
\blacken\path(4463.000,351.000)(4583.000,381.000)(4463.000,411.000)(4463.000,351.000)
\put(188,291){\makebox(0,0)[lb]{\scriptsize $d\xbold_t = \fbold(\xbold_t,\theta,$~\textcolor{red}{\ul{$u_t$}}$~) dt~+$}}
\put(600,100){\makebox(0,0)[lb]{\scriptsize $\Sigma(\xbold_t)^{1/2} dW_t$}}
\put(653,536){\makebox(0,0)[lb]{\smash{{\SetFigFont{10}{12.0}{\rmdefault}{\mddefault}{\updefault}{\bf System}}}}}
\path(2363,711)(4088,711)(4088,111)
	(2363,111)(2363,711)
\put(2660,497){\makebox(0,0)[lb]{\footnotesize {\bf State Estimator}}}
\put(2580,272){\makebox(0,0)[lb]{\scriptsize noisy observations~$\longrightarrow \hat{\xbold}_t$}}
\path(4583,681)(5558,681)(5558,111)
	(4583,111)(4583,681)
\put(4723,470){\makebox(0,0)[lb]{\smash{{\SetFigFont{10}{12.0}{\rmdefault}{\mddefault}{\updefault}{\bf Controller}}}}}
\put(4708,245){\makebox(0,0)[lb]{\scriptsize \textcolor{red}{\ul{$u_t = F(\hat{\xbold}_t,t)$}}}}
\put(2844.314,-3395.625){\arc{9323.250}{4.2769}{5.2121}}
\blacken\path(975.086,907.644)(878.000,831.000)(999.732,852.940)(975.086,907.644)
\path(1898,381)(2363,381)
\blacken\path(2243.000,351.000)(2363.000,381.000)(2243.000,411.000)(2243.000,351.000)
\put(915,396){\ellipse{1814}{780}}
\end{picture}
}
  \caption{Schematic representation of overall strategy.  A physical or
    biological system is modeled as a diffusion process with unknown
    parameters, and experiments are to be conducted to estimate these parameters.  Using available data
    up to time $t$, an estimate $\hat{\xbold}_t$ of the state of the
    system at time $t$ is estimated.  This is used to compute (in real
    time) a time-dependent control signal $u_t$ (underlined above),
    which drives the system toward states that are maximally-informative
    for the parameters in question.  The protocol, or {\em control
      policy}, is represented by the function $F(\xbold,t)$, which can
    be computed off-line.  In contrast to this ``closed-loop''
      set-up, in ``open-loop'' control schemes, controllers do not
      receive and cannot adapt to real-time state estimates.}
  \label{fig:schematic}
\end{figure}

As a concrete example, consider a typical {\em in vitro} neuron
experiment: an electrode is attached to an isolated neuron, through
which the experimenter can inject a time-dependent current $I_t$ and
measure the resulting membrane voltage $v_t$.  By injecting a
sufficiently large current, the experimenter can elicit a rapid
electrical pulse, or ``spike,'' from the neuron.  These are
  generated by the flow of ionic currents across voltage-sensitive
  membrane ion channels.
A standard model for spike generation in neurons is the Morris-Lecar
model \citep{MorrisLecar81}, which has the form
\begin{equation}
  \label{eq:ML}
  \begin{array}{rl}
    C_m dv_t =& \big[I_t - g_K~ w_t~(v_t-E_K) -
      g_{Ca}~m_\infty(v_t)~(v_t-E_{Ca})\\[1ex]
      &~~~~~- g_l~(v_t-E_l)\big]~dt +~\beta_vdW_1 \\[2ex]
    dw_t =& \frac{\phi}{\tau_w(v_t)}~(w_\infty(v_t) - w_t)~dt +
    \beta_w~\gamma(v_t,w_t)~dW_2\\
  \end{array}
\end{equation}
The auxiliary functions $w_\infty$, $\tau_w$, $m_\infty$~, and
$\gamma$ are given in Appendix~\ref{app:ml}.
In these equations, $I_t$ represents the externally applied, time-dependent current,
which we treat as the control variable; the variable $v_t$ is the
membrane voltage.  The first equation expresses Kirchoff's current
conservation law, and the constants $E_K$, $E_{Ca}$ and $E_l$ are
so-called ``reversal potentials'' associated with the potassium,
calcium, and leakage currents, respectively and represent the
equilibrium voltage that would be attained if the corresponding current
were the only one present.  The quantities $g_K w_t$, $g_{Ca}
m_\infty(v_t)$, and $g_l$ are the corresponding conductances; when
divided by the membrane capacitance $C_m$, they give the equilibration
rate associated with each type of current \citep{termantrout}.  The
dimensionless variable $w_t$ is a ``gating variable.'' It takes value in
$[0,1]$, and describes the fraction of membrane ion channels that are
open at any time.  The Brownian motions $dW_1$ and $dW_2$ model membrane voltage
fluctuations and ion channel noise, respectively.  (This model is
discussed in more detail in Sect.~\ref{sec:neuro}.)

In typical experiments, the current $I_t$ is a relatively simple
function of $t$, e.g., a step, square pulse, or a sinusoid, and the
amplitude of $I_t$ is adjusted so that a spike is triggered.  We seek
instead to set $I_t$ {\em in real-time}\footnote{This can potentially be
  accomplished in ``dynamic clamp'' experiments, in which the recorded
  voltages are fed directly into a computer, which also sets the
  injected current.  } according to the measured voltage trace up to
that time, i.e., $\{v_s:0\leq s\leq t\}$, and a
pre-computed ``control policy.''  The latter is designed to optimize the
Fisher Information gained about a parameter of interest, e.g., $g_{Ca}$ or
$C_m$, so that a precise estimate can be obtained more quickly.

In this paper, we propose a general, two-step strategy for designing
optimal control policies for such dynamic experiments.
First, assuming the experimenter can observe the full state vector
$\xbold_t$ for all $t$, we show that there exists a function $F$ such
that if the experimenter sets the control parameter to be $u_t =
F(\xbold_t,t)$ at all times $t$, then the Fisher information of the
unknown parameters will be maximized.  We call this the {\em full
  observation} case.  The optimal ``control policy'' $F$ in this case
can be precomputed off-line using stochastic optimal control techniques.
Second, in the more common {\em partial observation}
case where one only has partial, infrequent, and/or
noisy measurements of the system state, we show that stochastic optimal
control can be usefully combined with standard filtering methods in a ``closed-loop''
  set-up(see Fig.~\ref{fig:schematic}).
For simplicity, we focus here on the estimation of a single
  parameter in the system, but extensions to linear combinations of such
  criteria, such as the trace of the information matrix, are easy.

We illustrate the methods on two examples: a double-well potential example
taken from statistical physics and
the neuron example mentioned above. Appendix \ref{sec:chemostat} also illustrates an ecological experiment,
in which models of the form \eqref{eq:sde} are used to describe
  relative population sizes in a contained glass tank or chemostat, and
  the rate at which resources are added to and removed from the system
  is the control signal.
Together, these model systems represent useful contrasts in the
  quality of data, system behavior, and level of stochastic variation in
  system dynamics.

In particular, the double-well example presents a scenario where our
method might be especially useful: in that example, the relevant
parameter is a potential barrier height controlling the frequency of
 ``barrior-crossing'' events. he barrier
  occupies a small volume of phase space, but has a significant impact
  on the dynamics.  In this
  system, a dynamic control policy can dramatically increase
barrier-crossing frequency and hence the quality of our parameter
estimates; we do not expct control schemes that do not make use of
  current state information, including existing optimal ``open-loop''
  controls (see below), to be as effective. The other examples show
that our method can have varying degrees of impact: the neuron model
represents a realistic application in which membrane voltage can be
measured nearly continuously and with very high precision, and the
signals exhibit small, but important, stochastic fluctuations on top of
regular (e.g., oscillatory) behavior. Our control policy modifies the
shape of these oscillations to improve estimation precision but
parameters can also be reasonably estimated with a well-chosen constant
input. In Appendix \ref{sec:chemostat}, data from chemostat experiments
can be gathered at only infrequent intervals, and are subject to
variation due to binomial sampling of the tank and counting error.
Their dynamics exhibit relatively high stochasticity but tend to an
over-all fixed level, meaning that for long-run experiments, our methods
only alter the fixed-point and are not substantially different from
constant controls, although they may make a difference in short-term
experiments.

We view the main purpose (and contribution) of the present paper as
being the presentation of a novel problem in experimental design and
frameworks for approaching it, and to explore dynamical conditions
  when our method might be most effective.  Accordingly, the systems
studied here represent the simplest possible models and experiments in
their scientific domain, i.e., they are all low-dimensional, involving
only one or two state variables.  The strategy proposed in this paper
can potentially be applied to systems with more than two state
variables, but direct application of the numerical methods employed in
this paper to systems with more than a few dimensions may be
prohibitively expensive.  Higher-dimensional systems thus require
substantially new numerical techniques and ideas; this is beyond the
scope of the present paper, and is the subject of on-going work.

\paragraph{Related work}
Recent statistical literature has given considerable attention to
providing methods for performing inference in mechanistic, non-linear
dynamic systems models, both those described by ordinary differential
equations \citep*{Brunel08,RamsayDE,GCC10,HuangWu08} and explicitly
stochastic models \citep*{Ionides06,AitShahalia08,Wilkinson06} along
with more general modeling concerns such as providing diagnostic methods
for goodness of fit and model improvement \citep{Hooker08,MullerYang10,HookerEllner13}.
However, little attention has been given to the problem of designing
experiments on dynamic systems so as to yield maximal information about
the parameters of interest.  In this paper, we use stochastic optimal
control techniques to construct dynamic, data-driven experimental
protocols, which can improve the accuracy of parameter estimates based
on dynamic experiments.

To our knowledge, this is the first time that experimental design has
been employed to adaptively select time-varying inputs for explicitly
stochastic, nonlinear systems for the purposes of improving statistical
inference.  Within the context of stochastic models, adaptation can be
expected to be particularly important; information about parameters is
typically maximized at particular parts of the state space and the
choice of $u_t$ that will maintain a system close to high information
regions will depend on the current state of the system and cannot be
determined prior to the experiment.  In related work, experimental
design has been considered for ordinary differential equation models in
\citet*{Bauer00} and the choice of subject and sampling times has been
investigated in \citet{WuDing02} for mixed-effects models with ordinary
differential equations. \citet{HugginsPaninski11} consider the optimal
placement of probes on dendritic trees following linear dynamics.  A
number of recent studies have examined the use of perturbations to
improve parameter estimation, with a view toward biochemical reaction
networks; see
\citet{Busetto09,Nandy2012,Zechner2012,Ruess2013}.
However, in these papers the sequence of perturbations is decided prior
to the experiment, and is not adapted to the observations except in a
sequence of discrete experiments, i.e., these methods employ a
  form of ``open-loop'' control (Fig.~\ref{fig:schematic}).  We argue
that when they are feasible, adaptive experiments, i.e.,
  ``closed-loop'' control, are particularly advantageous.  They are
provably optimal among the class of all experiments. Because they are
based on the system's current state, they are likely to be particularly
helpful relative to non-adaptive states for systems with complex
dynamics that cannot be forecast with much precision.

The techniques applied in this paper rely on dynamic programming and
resemble methods in sequential designs, particularly for nonlinear
regression problems and item response theory. See
\citet{ChaudhuryMykland93} for likelihood-based methods or
\citet{ChalonerVerdinelli95} for a review of Bayesian experimental
design methods, \citet{Berger94} in item response theory and
\citet{BartroffLai10} for recent developments. Our problem, however, is
quite distinct from the sequential choice of treatments for independent
samples that is generally considered. In this paper, dynamic programming
is employed to account for the dependence of the future of the
stochastic process on the current choice of the control variable rather
than its effect on future design choices via an updated estimate (or
posterior distribution) for the parameters of interest. Indeed, unlike
most sequential design methods we will not attempt to estimate
parameters during the course of the experiment.  In
some contexts this can certainly be incorporated
  in our framework if so desired (see \citet{ThorbergssonHooker} for
  techniques for doing this in discrete systems), while in others ---
e.g., single neuron experiments --- this would
be computationally challenging given the
time-scales involved.  Where the optimal design depends on
the parameter, we propose averaging the Fisher Information over a prior,
corresponding to the use of a quadratic loss function in
\citet{ChalonerLarntz89}.


\paragraph{Organization}
In Sect.~\ref{sec:ControlTheory}, we provide a precise formulation of
the dynamic experimental design problem and outline our solution
strategy for ``full observation'' problems where we can observe the
state vector continuously.  Sect.~\ref{sec:Filtering} discusses our
approximate strategy for solving partial observation problems.  An
illustrative example, taken from statistical physics, is studied in
Sect.~\ref{sect:kramers}.  A detailed study of the Morris-Lecar neuron
model   is
presented in Sect.~\ref{sec:neuro}.
Some final remarks and directions for
further research are detailed in Sect.~\ref{sec:Conclusion}.
Appendices provide details of the numerical implementation of our algorithms and a
further study of chemostat models.
 We have
implemented all the algorithms described in this paper in R, and used
the implementation to compute the examples.  The source code is
available as an online appendix.



\section{Maximizing Fisher Information in Diffusion Processes} \label{sec:ControlTheory}

In this section we demonstrate that when the full state vector of the
system is continuously observable, the problem of designing optimal
dynamic inputs for parameter estimation can be cast as a problem in control theory.
Once this is done, we can follow well-developed methods.  A brief review
of some such methods is provided here (see also Appendix \ref{app:details});
interested readers are referred to \citet{Kushner00} for further
details.

The solution of the ``full observation'' problem described in this
section will form the basis for the more realistic but challenging
``partial observation'' problem, which we discuss in
Sect.~\ref{sec:Filtering}.

\subsection{Problem formulation and general solution strategy}
\label{sect:Full Observations}

Consider the multivariate diffusion process \eqref{eq:sde}.  Our goal is
to estimate $\theta$ using observations of $\xbold_t$ for $t\in[0,T]$,
up to some final time $T$.  At each moment in time, we assume the
experimenter can adjust the control $u_t$ using all available
information from the past, i.e., $\xbold_s$ for $s<t$.  Our problem is
to find an optimal control policy, i.e., a procedure for choosing a
control value $u_t$ based on past observations such that the resulting
estimator of $\theta$ is ``optimal,'' which we interpret here as meaning
``minimum asymptotic variance.''  Other criteria for experimental design
could be applied; \citet{Busetto09}, for example, maximizes the expected
divergence between prior and posterior for the purpose of model
selection \citep[see also][]{ChalonerLarntz89}.  Some (though not all) of these can be put in the
  same dynamic programing framework that we employ here.



To begin, recall that the likelihood of $\theta$ based on a single
realization or sample path of the diffusion process $\xbold_t$ in \eqref{eq:sde}
is given by
\begin{align*}
  l(\theta|\xbold) &= \frac{1}{2}\int_0^T
  \fbold(\xbold_t,\theta,u_t)^T\cdot
  \Sigma^{-1}\big(\xbold_t\big)\cdot\fbold(\xbold_t,\theta,u_t)~dt \\
  &- \int_0^T \fbold(\xbold_t,\theta,u_t)\cdot\Sigma^{-1}\big(\xbold_t\big)\cdot
  d\xbold_t
\end{align*}
\citep[see, e.g.,][]{Rao99}.  Since we are interested in choosing $u_t$
to minimize the asymptotic variance of the parameter estimate, we should
maximize the associated Fisher information
\begin{equation} \label{fofi}
I(\theta,u) = E \int_0^T \left|\left| \frac{d}{d\theta} \fbold(\xbold_t,\theta,u_t)
\right|\right|^2_{\Sigma(\xbold_t)} dt
\end{equation}
with $||\zbold||_{\Sigma}^2 = \zbold^T \Sigma^{-1} \zbold$.
Note that the Fisher Information \eqref{fofi} --- and hence the optimal
control policy --- in general depends on the parameter $\theta$ which
is, of course, unknown.  This can be handled by assuming a prior
distribution for $\theta$; see Sect.~\ref{sect:Parameter Dependence and
  Priors}.  For the discussion at hand, we assume that a reasonably good
initial guess of $\theta$ is available, which we use in computing
\eqref{fofi}. In this paper we focus on parameters that appear only in
the drift term $\fbold$, but note that similar expressions -- also
represented as integrals over the duration of the experiment -- can be
obtained for parameters that appear in $\Sigma$. However, unless
$\Sigma$ also depends on the state $\xbold_t$, the use of a control
policy that changes only the drift may not change the Fisher
Information for its parameters.

{\em A priori}, the control $u_t$ can be any function of the past
history $(\xbold_s:s<t)$ and $t$, i.e., it is a stochastic process
adapted to the filtration generated by $\xbold_t~.$ Note also that the
choice of a control at time $t$ affects the future cost, and hence
subsequent choices of the control.  Because of the form of \eqref{fofi}
as an integral over time and the Marcovian dynamics of the system, the
optimal control $u_t$ only depends on the current state $\xbold_t$
(and not its past).  That is, there is a function
$F:\R^d\times[0,T_{final}]\to{\cal U}$, where $\R^d$ is the state space,
$T_{final}$ is the duration of the experiment, and ${\cal U}$ is the set
of permissible control values, such that $u_t = F(\xbold_t,t)$ gives the
optimal control policy. Hereafter we will refer to this function $F$ as
the {\em optimal control policy}.



Computing the optimal control policy given an equation of the
form~(\ref{eq:sde}) is a standard problem in stochastic optimal control.
Here, we follow an approach due to Kushner (see, e.g.,
\citet{Kushner71,Kushner00,Bensoussan} for details), which consists of
two main steps: (i) a finite difference approximation is used to
discretize the problem in both space and time, thus reducing the problem
to one involving a finite-state Markov chain; and (ii) using the
well-known dynamic programming algorithm, construct an optimal control
policy for the discretized problem.  For conceptual and notational
simplicity, we review the basic dynamic programming idea --- applicable
to both continuous and discrete state spaces --- in the context of a
time-discretized version of the diffusion process, deferring details of
the spatial discretization to Appendix~\ref{app:details}.

\paragraph{Dynamic programming}
Let us approximate the diffusion process~(\ref{eq:sde}) at discrete
times $t_i = i \dt$, $i = 1,\ldots,T$ via the Euler-Maruyama scheme
\citep{kloeden1992numerical}, yielding
\begin{equation}
\xbold_{i+1} = \xbold_i + \fbold(\xbold_i,\theta,u_i)~\dt +
\sqrt{\dt}~\Sigma^{1/2}(\xbold_i) \epsilonbold_i
\label{eq:euler}
\end{equation}
with the $\epsilonbold_i$ independent vectors of independent standard
normal random variables, and approximate the Fisher Information by
\begin{equation}
  \widehat{FI}(\theta) = \sum_{i=1}^TE\left[ \left|\left|
    \frac{d}{d\theta} \fbold(\Xbold_i,\theta,u_i)
    \right|\right|^2_{\Sigma(\Xbold_i)}\right]~\dt~.
  \label{eq:fi}
\end{equation}
Note that in this section, we use $\Xbold$ to refer to
  non-realized random variables over which we take expectations, and use
  $\xbold$ to refer to specific realizations and deterministic values.

Our goal is now to choose $u_i$ at each step to maximize
$\widehat{FI}(\theta)~$ over all possible choices of $\ubold = (u_0,
u_1, \cdots, u_T)$, with the requirement that (i) $u_i\in{\cal U}$ for
all $i$, and (ii) $u_i$ is a function of $\xbold_j$ for $j<i$, i.e.,
$\ubold$ is adapted to the filtration generated by
$(\Xbold_0,\cdots,\Xbold_T)$.
%
%
To this end, we define for each timestep $i$ the {\em Fisher Information
  To Go} (FITG)
\begin{displaymath}
\widehat{FI}_i(\theta,\xbold) = \sup_{{\ubold}}\left( E_{\mbox{\scriptsize $\xbold$},i}\left[~\sum_{j=i}^T \left|\left|\frac{d}{d\theta}
  \fbold(\Xbold_j,\theta,u_j)\right|\right|^2_{\Sigma(\mbox{\scriptsize$\xbold$})}
  ~\right]\right)~,
\end{displaymath}
where $E_{\mbox{\scriptsize$\xbold$},i}(\cdot)$ denotes the conditional
expectation over all $\Xbold_j~,j>i$, given $\Xbold_i=\xbold$, and the
supremum is taken over all controls $\ubold = (u_i, u_{i+1},\cdots,
u_T)$ adapted to $(\xbold, \Xbold_{i+1}, \cdots, \Xbold_T)$.  The FITG
is the maximal expected FI, given that the we start at step $i$ in state
$\Xbold_i = \xbold$~; the optimal Fisher information
  $\widehat{FI}(\theta)$ in Eq.~(\ref{eq:fi}) is given by
  $\widehat{FI}_0(\theta,\Xbold_0)~.$

Dynamic programming is based on the observation that we can rewrite the
FITG at step $i$ recursively:
\begin{equation}
\widehat{FI}_i(\theta,\xbold) = \max_{u\in{\cal U}}\left( E_{\mbox{\scriptsize $\xbold$},i,u}\left[
  \widehat{FI}_{i+1}(\theta,\Xbold_{i+1}) + \left|\left|\frac{d}{d\theta}
  \fbold(\xbold,\theta,u)\right|\right|^2_{\Sigma(\mbox{\scriptsize$\xbold$})}
  \right]\right)~,
\label{fitg}
\end{equation}
where $E_{\mbox{\scriptsize$\xbold$},i,u}(\cdot)$ is the conditional
expectation given $\Xbold_i=\xbold$ and that we choose the control value
$u$ at step $i$~.  (Note that choice of $u$ affects not only the second
term inside the expectation, but also the first term, since choice of
$u$ affects $\Xbold_{i+1}$.)  This means that if we start at the {\em
  final} step $i=T$ and work progressively {\em backwards} in time, then
at each step $i$, the FITG $\widehat{FI}_{i+1}(\theta,\Xbold_{i+1})$
will already be computed.  Thus, the problem reduces to a
sequence of optimizations over ${\cal U}$, which is generally
straightforward to do (see below).  The resulting optimal control policy
$F$ is
\begin{equation}
  F(\xbold, i\dt)
  = \argmax_{u\in{\cal U}}\left( E_{\mbox{\scriptsize$\xbold$},i,u}\left[
    \widehat{FI}_{i+1}(\theta,\Xbold_{i+1}) + \left|\left|\frac{d}{d\theta}
    \fbold(\xbold,\theta,u)\right|\right|^2_{\Sigma(\mbox{\scriptsize $\xbold$})}
    \right]\right)~.
  \label{optcontrol}
\end{equation}
To compute $F$ on a computer, we can discretize the state space and
tabulate $F$ on a discrete grid of $\xbold$'s.
%
%
In this way, dynamic programming allows us to construct a good
approximation of the control policy $F$ for each time step $i$.


\subheading{Optimization over ${\cal U}$.}  The optimization of $u_i$ in
\eqref{optcontrol} may be done in a number of ways, depending on the
allowed set of control values ${\cal U}$~.  For example, if ${\cal U}$
is an interval or an open region in $\R^n,$ derivative-based
optimization techniques like Newton-Raphson may be useful.  In the
present paper, we assume the control values are drawn from a relatively
small finite set ${\cal U}$;
the optimization in Eq.~(\ref{optcontrol}) is thus done by picking the
maximum over this finite set via exhaustive search.

In more general situations than what we consider in this
paper, one may wish to allow ${\cal U}$ to be an unbounded set such as
$\R^k~.$ In these situations, the FITG may be maximized at very large or
even infinite values of $u_t$~.
%
%
As an example, consider the univariate Ornstein-Uhlenbeck process with
additive control
\begin{equation} \label{eq:ou}
dx = \left(-\beta x + u\right) dt + \sigma dW_t
\end{equation}
approximated via discrete time-steps as in \eqref{eq:euler}; at step $T-2$ the FITG for $\beta$ is $\xbold_{T-1}^2$ leading to
\[
u_{T-2} = \argmax_u E_{\xbold_{T-1}|\xbold_{T-2},u_{T-2}} \xbold_{T-1}^2 = \left( \left(1 -\dt \beta\right) \xbold_{T-2} +
\dt u_{T-2}\right)^2 + \dt \sigma^2
\]
which is maximized for $u = \pm \infty$.

Aside from the question of mathematical well-posedness of the
optimization problem~(\ref{optcontrol}), large values of $|u_t|$
constitutes an important practical issue: in experimental set-ups, $u_t$
may correspond to, say, an electrical current, and allowing $|u_t|$ to
become too large may have undesirable consequences.  An alternative to
restricting to a finite set of values is to allow continuous
  values of $u$ but add a penalty to the Fisher Information, so that
\eqref{optcontrol} has a finite, well-defined solution.

Throughout the discussion above, we have discretized time but assume
that expectations with respect to $\xbold_t$ can be calculated.  To make
dynamic programming practical, we further discretize the state space of
our system (using the methods of \citet{Kushner71}) to obtain a
finite-state Markov-chain approximation, to which dynamic programming
can be readily applied.  As the details of this implementation are
somewhat cumbersome to describe and not directly relevant to the
remainder of the paper, we refer interested readers to Appendix
\ref{app:details}.

\subsection{Parameter dependence and priors}
\label{sect:Parameter Dependence and Priors}

Throughout the discussion above, we have constructed the Fisher
Information $I(\theta,u)$ and the resulting control policy $u(\xbold,t)$
assuming a specific $\theta$.  In general, $\theta$ will not be known
--- there would otherwise be little point in the experiment --- and both
Fisher Information and the optimal control policy may depend on
its value.

We address this issue by constructing a prior $\pi(\theta)$ over
plausible values of $\theta$ and maximizing $E_{\theta}(I(\theta,u))$.
The choice of this prior is important: the dynamics of $\xbold_t$ may
depend on the value of $\theta$, and the computed control
policy may be ineffective if the value of $\theta$ assumed in computing
the optimal control policy is very different from the true value.  This
averaging is easy to implement numerically: a grid is chosen over the
relevant region of parameter space, then the FI \eqref{fofi} is averaged
over this grid of $\theta$ (a weighted average can be used to implement
a non-uniform prior).  It is easy to check that all the algorithms
described in Sect.~\ref{sec:ControlTheory}, as well as those described
below, apply in a straightforward way in this setting.

Averaging Fisher Information over a prior in this way corresponds to the
use of a quadratic loss function in a Bayesian design setting as in
\citet{ChalonerLarntz89}. We have employed this here as we wish to
minimize the variance of the estimated parameter. A Bayesian D-optimal
design corresponding to averaging log Fisher Information (or its log
determinant if $\theta$ is multivariate) can be employed following the
same methods.

We note that while averaging the objective over possible values of the
parameter is a natural procedure in some situations, it may not always
be appropriate.  When an estimate of parameters can be obtained and
updated as the experiment progresses, it may be preferable to use the
control policy associated with using these on-line estimates, e.g., by
pre-computing a set of control policies (for a grid of parameter values)
and use the current best parameter estimate to choose a control policy.
Alternatively, as the experiment progresses, Fisher Information could be
averaged over a posterior for the parameters.  These ideas are explored
in \citet{ThorbergssonHooker} in the context of discrete-state Markov
models; such on-line strategies are more complicated to implement for
the types of diffusion processes discussed here, and will be examined in
a future publication.

\section{Partial Observations}
\label{sec:Filtering}

The algorithms discussed in Sect.~\ref{sec:ControlTheory} are
appropriate for a model in which the entire state vector $\xbold_t$ is
observed essentially continuously in time, without error.  This is
rarely the case in practice.  In neural recording models, only membrane
voltages can be measured, leaving $w_t$ as latent, although voltage
measurements have both high accuracy and high frequency. In the
chemostat system described in Appendix~\ref{sec:chemostat},
%
%
algae are measured by placing a slide with a sample from the ecology in
a particle counter and are thus subject to sampling and counting errors
and can be taken at most every few hours. Nitrogen can also be measured
but with less accuracy and greater expense.


Generally, not all state variables will be measured in most
applications, and measurements will be taken at discrete times and are
will likely be corrupted by measurement errors. We denote both
conditions by the term {\em partial observations}.  The observation
process causes two problems for the strategy outlined in
Sect.~\ref{sect:Full Observations}: first, since not all state variables
are being measured, the Fisher information in Eq.~(\ref{eq:fi}) is no
longer the correct expression for the asymptotic estimator variance; and
second, the dynamic programming methodology outlined in
Sect.~\ref{sec:ControlTheory} is no longer applicable.

These difficulties associated with partially observed systems are of a
fundamental nature: projections of Markov processes are typically not
Markovian, and the Markov property is essential to the dynamic
programming algorithm (see Sect.~\ref{sect:Full Observations} and
references). While one can derive an explicit expression for the Fisher
information of partially observed diffusion processes, the functional no
longer has the form of a time-integral of a function of the state vector
(see \cite{Louis82}), making it much more difficult to work with. Even
when algebraic expressions for the Fisher Information are available,
either in specific cases (linear diffusions with continuously-observed
stochastic noise, also discrete-state systems) or by approximation
\citep{Komorowski2011,Komorowski2012}, a straightforward application of
dynamic program cannot be employed directly. See
\citet{ThorbergssonHooker} for approximations in discrete-state systems.

In this section, we propose a simple approximation strategy aimed at
overcoming these difficulties.  We also provide some theoretical
justification for the approximation strategy in the context of linear
systems.  We assume throughout this section that the dynamics have been
time-discretized (see Sect.~\ref{sect:Full Observations}) to yield a
sequence of state vectors $\xbold_0, \xbold_1, \cdots, \xbold_T$, and
that every $\tau$ steps a noisy observation is made, yielding the
observation vectors $\Ybold = (\ybold_1,\ldots,\ybold_{n})$~, where
$n=T/\tau$ is the number of observations.  Furthermore, we assume that a
noise model $p(\ybold|\xbold)$ is given.  The assumption of
strictly time-periodic observations is for simplicity only and
can be easily relaxed.

\subsection{Filtering and estimation for partially observed systems}
\label{sect:Filtering and estimation for partially observed systems}

Our strategy for dealing with partially observed systems entails the
following steps:
\begin{enumerate}

\item[(i)] We solve the full observation problem by finding the control
  policy $F$ that minimizes the full-observation Fisher Information
  (FOFI) given by \eqref{fofi}.

\item[(ii)] During the experiment, filtering techniques are applied to
  provide real-time estimates of the system state, which are then
  plugged into $F$ to obtain a control value.

\end{enumerate}
While this strategy is not expected to be optimal in general because
the control values are determined by the current state rather than the
entire past history, we argue that it nevertheless provides an
efficient control strategy for problems where accurate filtering is
feasible, i.e., when the conditional variation of the process given
observations of it is small.

Step (i) above follows exactly the methodology from
Sect.~\ref{sec:ControlTheory}.  The main difference between the partial
and full observation cases is the need for filtering.  For conceptual
and implementation simplicity, in this paper we use a version of a {\it
  particle filter} \citep[see, e.g.,][]{liu}.  These allow us to
  both produce a filter distribution for the
  $\xbold_t|\ybold_1,\ldots,\ybold_t$ and also to calculate a likelihood
  for each candidate value of $\theta$. This allows us to employ a
  maximum likelihood estimate of $\theta$ in our simulations. The
  details of our particle filter and likelihood maximization algorithm
  are given in Appendix \ref{app:filter}.

\subsection{Approximation Methods and Linear Systems}

We now provide a partial theoretical justification of our approximation
strategy for systems in which the observations are very informative
about $\xbold_t$.  This requires us to both examine the utility of the
full observation Fisher Information as well as the use of a filtered
estimate of $\xbold_t$ within the control policy.

Beginning with the approximation to Fisher information,
within a partially observed system the complete data log likelihood can
be written as
\[
l(\theta|\xbold,\Ybold) = \sum_{i=1}^n \log p(\ybold_i|\xbold_i) + l(\theta|\xbold)
\]
assuming that $\theta$ does not appear in the observation process. From the
formulation for the observed Fisher Information
\begin{align*}
I(\theta|\Ybold) & = E_{\xbold|\Ybold} \frac{d^2}{d \theta^2} l(\theta|\xbold,\Ybold)
 - \mbox{var}_{\xbold|\Ybold} l(\theta|\xbold,\Ybold)
\end{align*}
the (expected) Fisher Information for the partially observed diffusion process can be
written as
\begin{align*}
I_Y(\theta,u) =& E_{\xbold} \int \left| \left|\frac{d}{d \theta} \fbold(\xbold_t,\theta,u_t)\right| \right|_{\Sigma(\xbold_t)}^2 dt \\&-
E_Y \mbox{var}_{\xbold|\Ybold} \left( \int \left[\frac{d}{d \theta} \fbold(\xbold_t,\theta,u_t)\right] \Sigma(\xbold_t)^{-1} \left(d\xbold_t - \fbold(\xbold_t,\theta,u_t)dt\right) \right)
\end{align*}
\citep[see][]{Louis82} where we observe
that the first term is the objective of our dynamic program. Taking a
Taylor expansion of $d\fbold/d\theta$ in $\xbold_t$ centered on $E
\xbold_t |\Ybold$, we observe that the second term shrinks with
$\mbox{var}(\xbold|\Ybold)$ assuming that $d\fbold/d\theta$ is
differentiable in $\xbold$.

The argument above indicates that, provided the conditional variance of
$\xbold$ given observations $Y$ is small, FOFI is a reasonable
approximation to the Fisher Information obtained from $Y$. Indeed, in
the limiting case that this variance goes to zero, we recover FOFI
exactly.

It remains necessary to demonstrate that employing the filtered estimate
within the control policy will approximately maximize FOFI. A
  full theoretical justificaiton, however, is not
  straightforward: for
linear diffusion processes, $d\xbold_t = (A\xbold_t + Bu_t)dt + dW_t$,
with a quadratic objective function $\int \xbold_t^T C(t) \xbold_t~dt$,
the Separation Theorem guarantees that this strategy is optimal
\citep[see][]{Kushner71}.
This needs not be the case for nonlinear systems or non-quadratic
objective functions.  Extensions of the separation theorem have been
developed in \citet{KilicaslanBanks09} based on successive
approximations of \eqref{eq:sde} in terms of a linear system with
time-varying parameters. Particle filter methods can also be employed to
average the future cost over the current distribution of the estimated
state variables \citep{ADST03}. Both of these schemes require re-running
the dynamic program at each time point, a strategy that will not always
be computationally feasible.

In contrast, the approach of using the filtered estimate of the state
allows the map from state variables to controls to be pre-computed. This
was employed in \citet{BotchuUngarala07} for optimal control, for
example, and we demonstrate here that it is
also effective for estimating parameters. If we consider the
limit of informative observation, that is in a sequence of systems in
which $\left|\left|\xbold^* - \xbold\right|\right| \rightarrow 0$ where
$\xbold^*$ is the filtered estimate, then the Fisher Information
resulting from this procedure can be shown to converge to its optimal
value under suitable regularity conditions. The details of the
conditions required will depend on context-specific factors such as
whether the control policy takes discrete or continuous values, and we
do not attempt to fill these in here. The behavior suggested here can be
seen in the comparative performance of results in the
Morris-Lecar and chemostat simulations examined here, as well as in
results in \citet{ThorbergssonHooker}.


\section{An illustrative example}
\label{sect:kramers}

\begin{figure}
  \begin{center}
    \begin{tabular}{ccc}
      \resizebox{1.8in}{1.25in}{\includegraphics[bb=0in 0in 4in 3in]{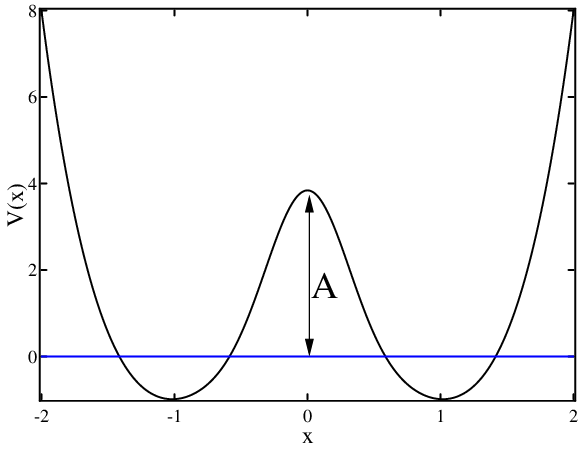}}&&
      \resizebox{2.2in}{1.25in}{\includegraphics[bb=0in 0in 4in 3in]{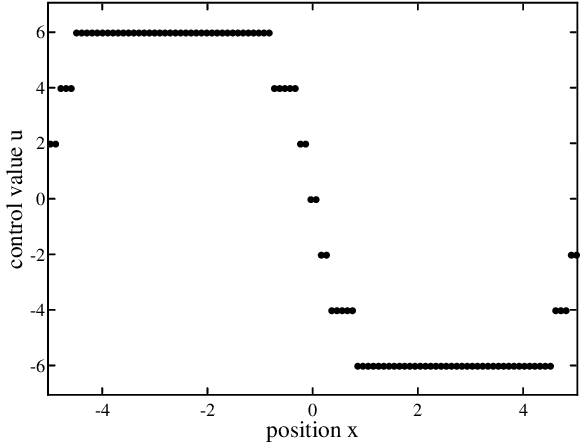}}\\
      (a) The potential $V(x)$ &&
      (b) Optimal control at $t=0$\\
    \end{tabular}\\[4ex]
    \resizebox{4.5in}{1.25in}{\includegraphics[bb=0in 0in 5in 2in]{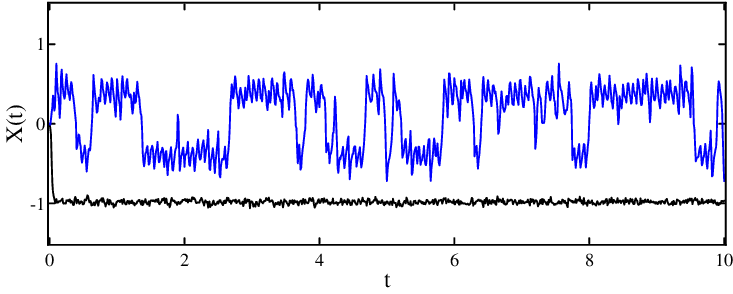}}\\
    (c) Controlled and uncontrolled sample paths
  \end{center}
  \caption{Particle in a double-well potential.  In (a), the potential
    (\ref{eq:double-well}) is shown.  Panel (b) shows a plot of the
    computed optimal control policy, as a function of position $x$, for
    $t=0$ and $T=30$.  Panel (c) compares the controlled (lighter, blue
    curve) and uncontrolled (darker, black curve) sample paths.}
  \label{fig:kramers-pot}
\end{figure}

To illustrate our method and its impact, we present a simulation study for a model in which we expect
our methods will yield significant improvements in parameter estimation. We consider a small particle trapped in a
double-well potential (see Fig.~\ref{fig:kramers-pot}(a)).  This is a
commonly-used model in statistical and chemical physics
\citep{van-kampen,hanggi1990reaction}; we use it here to illustrate our
methodology, and to indicate when our methodology may be
  particularly effective.

The governing equation is
\begin{equation}
  dx_t = -V'_A(x_t) + u_t + \sigma~dW_t~,
  \label{eq:langevin}
\end{equation}
where $x_t\in\R$ is the position of the particle and the potential $V$
is given by
\begin{equation}
  V(x) = x^4 - 2x^2 + A e^{-(x/w)^2/2}~.
  \label{eq:double-well}
\end{equation}
(See Fig.~\ref{fig:kramers-pot}(a).)  Here, we study the model with
$w=0.3$, $\sigma=0.1$, and $A\approx 3.8$. We selected the last value randomly
  picked from an interval centered around $A=4$ in order to avoid cherry-picking
  favorable parameter values.  In this regime, the
unforced system (i.e., with $u_t\equiv 0$) exhibits dynamics on two
separate timescales: on the shorter timescale, the particle position
fluctuates about the bottom of one of the wells due to thermal
fluctuation; on longer timescales, on the order of $e^{2(\Delta
  V)/\sigma^2}$ where $\Delta V = A+1$ is the depth of the well, the
system jumps between the two potential wells
\citep{van-kampen,freidlin-wentzell}.  A sample path is shown in
Fig.~\ref{fig:kramers-pot}(c) (bottom, dark black curve).

Models of this type are paradigms for physical systems with multiple
metastable states, and calculating transition rates between metastable
states (due to large deviations in the driving white noise process) are
of interest in, e.g., in reaction rate theory.  By extension, the heights
of relevant potential barriers are also of interest.  Our goal here is
to estimate the barrier height $A$ in Eq.~(\ref{eq:double-well}).

For this 1D model, it is straightforward to solve the optimal control
problem.  Specificially, we discretize the interval $(-5,5)$ (with high
probability the position $x_t$ will stay within this range) by 100 grid
points.  The control values $u_t$ are drawn from the finite set $\{0,
\pm2, \pm4, \cdots, \pm10\}$.  We also assume a uniform prior for $A$ on
the interval $[2,5]$, which we discretize into 10 parameter values.  The
diffusion process~(\ref{eq:langevin}) is discretized in time with a
timestep of $\dt=0.01$; this is sufficiently small to satisfy the
criteria set forth in Sect.~\ref{app:details}.  An optimal control
policy is computed over the time interval $t\in[0,T]$ for various values
of $T$.  The computed control for $t\ll T$ stabilize rather quickly as
$T$ increases; the control at $t=0$ is shown in
Fig.~\ref{fig:kramers-pot}(b).  Not surprisingly, the control encourages
more frequent jumps by pushing left when the particle is in the right
well, and vice versa.

To see the control policy ``in action,'' we carry out simulations of the
controlled diffusion proceess for $T=4$ and for $T=30$.  For each choice
of $T$, we carry out 256 independent trials, and use the result to
estimate the barrier height $A$.  The diffusion process is observed
every 0.25 units of time; at observation times the system state is
estimated, and the control value updated.  The observations are assumed
to have additive, Gaussian observation noise of standard deviation 0.05.
Because of the observation noise, even though this is a 1D model (and so
the full state ``vector'' is observed), filtering is still needed.
Here, we use a particle filter with $10,000$ particles; far fewer
particles would have sufficed for the controlled process, but the
uncontrolled process needed more particles to obtain a reasonable
parameter estimate.  A sample trajectory subjected to the optimal
control is shown in Fig.~\ref{fig:kramers-pot}(c) (lighter / blue
curve).

\begin{table}
  \begin{center}
    \begin{tabular}{c|c|c|c|c|c|c|c}
      Duration & Control & N & In range & Mean & Bias & Std.~Dev. & Std.~Dev.~Err. \\\hline\hline
      4 & Dynamic & 256 & 100\% & 4.233 & 0.3933 & 0.05947 & 0.002628 \\
      4 & 0 & 256 & 98\% & 4.242 & 0.4016 & 0.3133 & 0.01385 \\
      30 & Dynamic & 256 & 100\% & 4.225 & 0.3854 & 0.02098 & 9.272e-4 \\
      30 & 0 & 256 & 99.6\% & 4.223 & 0.3832 & 0.2888 & 0.01277 \\
    \end{tabular}\\[2ex]
    (a) Continuous noise-free observations\\[4ex]

    \begin{tabular}{c|c|c|c|c|c|c|c}
      Duration & Control & N & In range & Mean & Bias & Std.~Dev. & Std.~Dev.~Err \\\hline\hline
      4 & Dynamic & 256 & 100\% & 4.225 & 0.3846 & 0.1094 & 0.004833 \\
      4 & 0 & 256 & 77\% & 4.202 & 0.3623 & 0.6048 & 0.02673 \\
      30 & Dynamic & 256 & 100\% & 4.197 & 0.357 & 0.04881 & 0.002157 \\
      30 & 0 & 256 & 71\% & 4.316 & 0.4763 & 0.5953 & 0.02631 \\
    \end{tabular}\\[2ex]
    (b) Infrequent, noisy observations

  \end{center}

  \caption{Comparison of estimates of the energy barrier height in the double well potential example using
    controlled and uncontrolled diffusions.   {\em in-range} = number of trials where
    estimator falls within uniform prior; {\em bias} = difference between
    the mean over tries and the true value; {\em std. dev.} = standard
    deviation of the estimator; and {\em std.~dev.~err.} = our estimate of
    the standard error of the standard deviation.
    Dynamic control shows a clear improvement in the precision of estimates.}


  \label{tab:kramers}
\end{table}

Table~\ref{tab:kramers} shows the results of these trials.  In
Table~\ref{tab:kramers}(a), results are shown for full observation
trials, in which the system is observed at every timestep, i.e.,
observation period = $\dt=0.01$~.  In addition to the controlled
diffusion process, we also computed $A$ estimates for the diffusion
process with a range of constant control values $u\in{\cal U}$, and
found that $u=0$ gives minimum-variance estimates among the control
values tested.  As can be seen, the standard deviation of the estimate
based on the controlled diffusion is significantly smaller for both
$T=4$ and $T=30$ --- for $T=4$ by a factor of $\approx5,$ and for $T=30$
by a factor of $\approx14.$ We note that there is a measurable bias in
the estimates from both dynamic and constant control regimes. In
practical experiments, this can be examined and corrected by conducting
a simulation; we note that the dynamic control policy does not appear to
affect the magnitude of this bias.

Table~\ref{tab:kramers}(b) shows the corresponding results for the case
of noisy-and-infrequent observations.  For $T=4$, we obtain a roughly
factor of 6 reduction in standard deviation, while for $T=30$, roughly a
factor of 11.  The reduction in the estimator variance is significant,
though as expected somewhat less than the full observation case, at
least over longer timescales.  More telling is the fraction of estimates
that were ``in-range'' (see the discussion of MLE in
Appendix \ref{app:filter}): the controlled diffusion process always
produced estimates that were within the prior parameter range, whereas
the uncontrolled diffusion produces a nontrivial number of parameter
estimates outside the range.

The reason that the optimal control strategy is particularly effective
in this model is that while the barrier height $A$ has a significant
impact on the dynamics of the system over long timescales, it only
impacts the dynamics in a small part of state space, and on shorter
timescales it has relatively little effect.  The optimal control policy
is able to drive the system into crossing the barrier much more
frequently, thereby yielding more information about $A$.  Note in
  order to know which way to drive the system to speed up
  barrier-crossing, the controller needs to know which state the
  particle is in.

We expect that, in general, our method will be particularly effective in
situations like this, where the information-rich region of phase space
is relatively small (in terms of how much time the uncontrolled
trajectory spends there), and the control is able to increase the
frequency of visits to these regions.


\section{Morris-Lecar (ML) Neuron Model}
\label{sec:neuro}

We now demonstrate the application of the optimal control
methodology to a more complex model with different dynamics,
with the goal of examining how the type of dynamics can affect the
efficacy of parameter estimation in the presence of dynamic control. Appendix \ref{sec:chemostat} presents a further exploration of a different system where our methods are less helpful.
 The Morris-Lecar (ML) neuron model mentioned in the introduction is often used to model membrane voltage
oscillations and other properties of neuronal dynamics; it describes how
membrane voltage and ion channels interact to generate electrical
impulses, or ``spikes,'' which is the primary means for neuronal
information transmission.  The ML model is planar, and hence more
amenable to analysis than higher-dimensional models like the
Hodgkin-Huxley equations \citep{HodgkinHuxley52}.  At the same time, it
faithfully captures certain important aspects of neuronal dynamics
\citep{ermentrout-rinzel}, making it a commonly-used model in
computational neuroscience.  It has a rich bifurcation structure, and
exhibits two dramatically different timescales.  See \citet{termantrout}
and \citet{ermentrout-rinzel} for a derivation and description of this
model and its behavior.

The ML model is given by \eqref{eq:ML}. To provide further details, the second
equation can be interpreted as the master equation for a two-state
Markov chain describing the opening and closing of ion channels.
The terms with $\beta_v$ and $\beta_w\gamma(v,w)$ are independent noise
terms modeling different sources of noise: the $\beta_v$ term models
voltage fluctuations, and is simply additive white noise.  The
$\beta_w\gamma(v,w)$ term models random fluctuations in the number of
open ion channels due to finite-size effects; here we use the function
\begin{equation}
  \gamma(v,w) = \sqrt{\frac{\varphi}{\tau_w(v)}
    \Big(w_\infty(v)\cdot(1-2w) + w\Big)},
\end{equation}
to scale the Wiener process. This function arises from an underlying
Markov chain model; see, e.g., \citet{kurtz71,kurtz81,smith}.

Here, we assume a typical experimental set-up in which an electrode (a
``dynamic clamp'') is attached to the neuron, through which an
experimenter can inject a current $I_t$ and measure the resulting
voltage $v_t$; the gating variable $w_t$ is not directly observable.
The electrode is usually attached directly to a computer, which records
the measured voltage trace and generates the time-dependent injected
current, making this type of experiment a natural candidate for our
method. Neuronal membrane voltages are measured with signal-to-noise
ratios of 1000 to 1 or more.  On the other hand, dynamical events of
interest can take place on timescales of milliseconds or less.  So for
this application it is vital that one pre-computes as much as possible.
These speed requirements may necessitate the use of computationally
cheaper --- but less accurate --- state estimation technique, e.g.,
ensemble Kalman filters.  Since we are interested evaluating the utility
of dynamic control via simulations, we continue to use particle filters
here.

\begin{figure}

  \begin{center}
    \begin{tabular}{cc}
      \includegraphics[bb=0in 0in 3.5in 3.5in,scale=0.6]{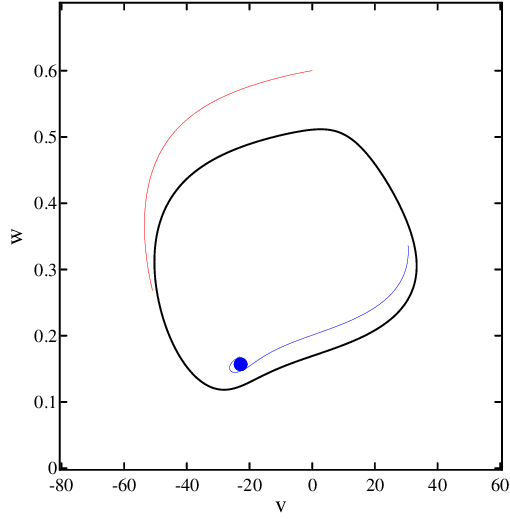}&
      \includegraphics[bb=0in 0in 3.5in 3.5in,scale=0.6]{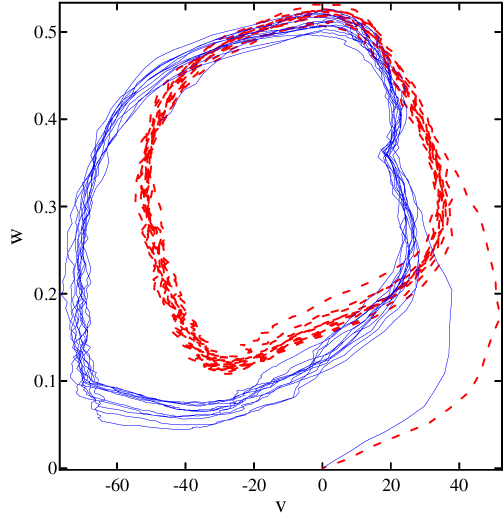}\\
      (a) Noiseless ML phase portrait & (b) Controlled and uncontrolled trajectories\\
    \end{tabular}
  \end{center}

  \caption{Morris-Lecar (ML) phase portrait of a stimulated neuron.  See \eqref{eq:ML}.  Time is measured
    in ms, voltage in mV, current in pA.  In (a), the phase
    portrait of the noiseless ML system is shown.  A stable limit cycle
    surrounds an unstable cycle, which in turn surrounds a sink.
    In (b), two trajectories for the noisy system are shown: the dotted
    curve is a trajectory without control, while the solid curve is a
    controlled trajectory.}
  \label{fig:ml-hopf-portrait}
\end{figure}

We focus on a parameter regime in which the noiseless system can switch
between having a globally-attracting fixed point and an unstable fixed point
surrounded by a stable limit cycle (Fig.~\ref{fig:ml-hopf-portrait}(a)
illustrates the latter).
The limit cycle represents a ``tonically'' (periodically) spiking
neuron; the effect of noise is to ``smear out'' this limit cycle.  The
precise parameters we use come from \citet{termantrout}; they are
summarized in Appendix~\ref{app:ml}.  The key parameter here is the
injected DC current; in Fig.~\ref{fig:ml-hopf-portrait}(a) this is
$I\equiv 100.0$.  As one decreases $I$ from this value, the system
undergoes a  {\em subcritical Hopf bifurcation:} the
unstable fixed point becomes stable, and at the same time an unstable
periodic orbit emerges around the fixed point \citep{termantrout}.
Post-bifurcation, the system is bistable: while tonic spiking continues
to be viable, a quiescent state (corresponding to the stable fixed
point) has emerged.  If we decrease $I$ even further, the stable and
unstable cycles collide in a saddle-node bifurcation, leaving behind a
single stable fixed point.

In this example, our control is $u_t = I_t / C_m.$ The control values
we use are $u\in\{0, 3.5, 5\},$ corresponding to $I\in\{0, 70, 100\}$
(we take $C_m=20$).  The three values of the injected current place the
system in the stable fixed point, bistable, and limit cycle regimes,
respectively.

\paragraph{Simulation results}
We have performed simulations in which we estimated the calcium
conductance constant $g_{Ca}$ from simulated data; this is chosen over
other parameters because it gives rise to a  more
  complex control policy (as we show below).  In detail, we assume a
flat prior for $g_{Ca}$ on the interval $[4,5]$; in all experiments
reported below, we fix a randomly-generated $g_{Ca}$ value of 4.415,
which we take to be the ``true'' value.  The optimal control is computed
using a timestep of $\approx 2$ ms, and we assume that measurements are
available at every time step.  State space is discretized by cutting the
region $[-80,80]\times[0,1]$ into $72\times72$ bins.  Unless otherwise
stated, ``constant control'' means $u_t\equiv 5$.
The value of $u=5$ was chosen as the best performing among several
  possible values of a constant control, which all exhibited
  significantly higher standard deviations.

\begin{figure}
  \begin{center}
    \begin{tabular}{cc}
     \includegraphics[bb=0in 0in 5in 4in,height=1.9in]{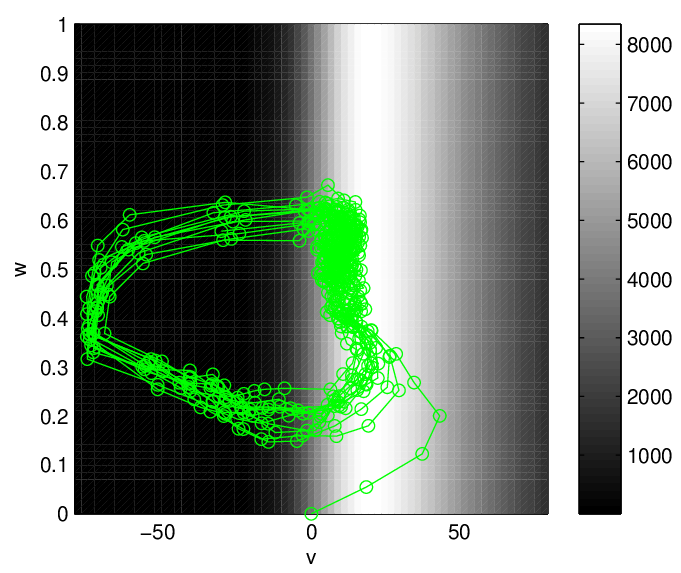} &
     \includegraphics[height=1.9in]{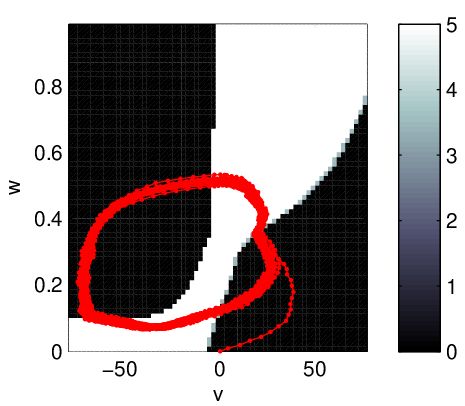} \\
      (a) & (b) \\
    \end{tabular}
  \end{center}
  \caption{Control policy for parameter $g_{Ca}$ in the Morris-Lecar
    example \eqref{eq:ML}.  In (a), the FI is shown behind a single
    controlled trajectory.  The lighter the region, the more informative
    that part of state space is about $g_{Ca}$.  In (b), the control
    policy is depicted behind a single controlled trajectory.  The
    control is adapted to the current state of the system in order to
    keep the trajectory in more informative regions.}
  \label{fig:ml-gca-control}
\end{figure}

Fig.~\ref{fig:ml-hopf-portrait}(b) shows examples of controlled and
uncontrolled (with $u=5$, or $I=100$) trajectories: the perturbations
are measurable, but not large.  Simulation results for estimates of
$g_{Ca}$ using trajectories of duration $T=1000$ ms are given in Table
\ref{mlsim:tab1}.

\begin{table}
  \begin{center}
    \begin{tabular}{c|c|l|l|l}
      Observation & Control & Mean & Bias & Std. Dev. \\\hline
      Full & Dynamic & 4.416 & 6.315e-4 & 0.01106 \\
      Full & 5.0 & 4.416 & 8.82e-4 & 0.01405 \\
      Noisy partial & Dynamic & 4.413 & -0.002062 & 0.01579 \\
      Noisy partial & 5.0 & 4.416 & 0.001458 & 0.01847 \\
    \end{tabular}
  \end{center}
  \caption{Simulation results for the Morris-Lecar example.     ``Full
    observation'' runs use exact information from the entire state-space
    trajectory, while ``Noisy partial'' runs correspond to only
    observing $v_t$ corrupted by observational noise.  Here, $T=1000$.}
  \label{mlsim:tab1}
\end{table}

 To better understand
why the dynamic control performs better, we examine the structure
of the control policy and Fisher information.
Fig.~\ref{fig:ml-gca-control}(a) shows the Fisher information function
for $g_{Ca}$, which is easily shown to be proportional to
\begin{displaymath}
  \big[m_\infty(v_t)~(v_t-E_{Ca})\big]^2~;
\end{displaymath}
the controlled trajectory is superposed.  From the geometry, one
  expects the optimal control should either find a way to keep the
  trajectory at a constant voltage or, failing that, try to increase the
  firing rate, so that trajectories cross the information-rich region as
  much as possible.  As can be seen, the controlled trajectory does
exactly that: starting at the resting voltage (around -80 mV), it jumps
toward 0 mV, runs along the ``information-rich'' strip around $v=20$
before dipping back toward the resting voltage.  In comparison, the
uncontrolled trajectory in Fig.~\ref{fig:ml-hopf-portrait}(b) appears to
spend less time in this region.  Fig.~\ref{fig:ml-gca-control}(b) shows
how the optimal control policy accomplishes this by suitably pushing the
trajectory at critical parts of its cycle, thus increasing the
  overall firing rate.


\begin{figure}
  \begin{center}
    \includegraphics*[height=1in]{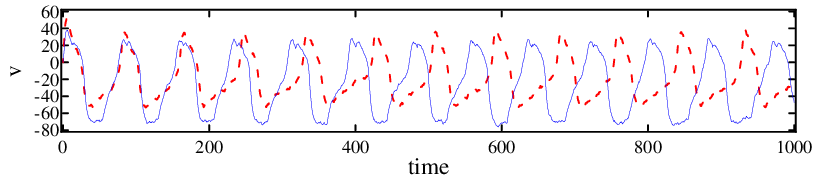}\\[1ex]
    \includegraphics[height=1in]{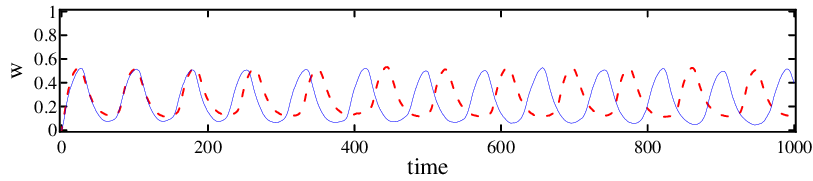}\\[-0.5ex]

  \end{center}
  \caption{Trajectories of the Morris-Lecar model under constant control (dashed) and optimal control (solid).  We see the optimal control has higher frequency oscillations which force the trajectories through the most informative state space more often.}
  \label{fig:ml-traces}
\end{figure}

To see more clearly the effect of the dynamic control, we plot time
traces of the controlled and uncontrolled trajectories in
Fig.~\ref{fig:ml-traces}.  As can be seen, the dynamic control not only
tries to push the trajectory into the information-rich region, it also
makes sure it spends more time there.

Some final remarks on the structure of the optimal control:
\begin{enumerate}

\item We note that the optimal control depends a great deal on the
  parameter being estimated.  If one were to try to estimate $C_m$, for
  example, the resulting control policy is essentially constant for most
  of phase space.

\item In this dynamic regime, with FI as in
  Fig.~\ref{fig:ml-gca-control}(a) (a unimodel function of $v$), it is
  natural to expect that increasing firing rate leads to an increase in
  the FI.  This can be accomplished by either dynamic control or by
  static control.  In more general situations, e.g., neuron models with
  additional currents and exhibiting dynamics on multiple timescales
  \citep{dayan2001theoretical}, we expect dynamic control to yield a
  greater gain in FI.  However, since such models usually entail $>3$
  degrees of freedom, they would be difficult to study using present
  numerical methods, and we leave the study of such models for future
  work.
\end{enumerate}


\section{Discussion} \label{sec:Conclusion}

%
%
%

There has been increasing interest in combining experimental data and
statistical methods with mechanistic dynamical models describing system
behavior. This paper adds to this literature in considering the problem
of designing inputs into such experiments that are directed at improving
the precision of parameter estimates that result from them. In
particular, we have demonstrated that in diffusion processes, maximizing
Fisher Information about a parameter can be cast as a problem of optimal
control and shown that using this strategy can substantially improve
parameter estimates.  In order to make control methods feasible when
systems are observed with noise or only some state variables are
observable, we employ the strategy of estimating the value of the state
on-line and using this within a pre-computed control policy.

We have demonstrated this approach on two
%
%
examples that showcase when this form of adaptive control is most likely
to be useful.  One situation when we expect significant benefit from
dynamic control is when visits to the information-rich regions of state
space are relatively rare, as in the double-well example.  As we have
seen, optimal control can be effective in increasing the frequency of
visits to information-rich regions.  In systems where trajectories
naturally return to information-rich regions in a recurrent fashion even
with static controls, our method may still yield modest benefits, but
the exact degree of information gain will depend on the specific dynamic
situation.

In systems with stable fixed points that are already informative about
the parameters of interest, dynamic control may not be better than simply choosing an optimal static control.
However, our methods also yield information about which constant control
is likely to be most useful; this is helpful in,
e.g., the chemostat experiments described in Appendix
\ref{sec:chemostat}, where measurements are relatively infrequent and
dynamic control is relatively easy to implement.
In systems such as \eqref{eq:ou} in which the location of fixed
points do not yield information about parameters, a control policy that
repeatedly returns the system to informative transient states could be
more advantageous.

There remain numerous unresolved problems and open areas in which these
methods can be extended.  Computational cost, both in speed and memory,
represent the largest limiting factor in employing these methods. In
particular, the storage and computation costs of the policy scale
exponentially with the number of state variables. Possible strategies
here involve the application of sparse grid strategies for discretizing
the state variables \citep{xiu} or -- more heuristically -- intensive
Monte Carlo methods to simulate the state variables forward combined
with machine learning methods to estimate control strategies.  It may
also be possible to pre-compute control policies only for
high-probability regions of the state space and employ techniques of
approximate dynamic programming \citep{powell}.

Further extensions involve alternative targets for the control policy,
including the Bayesian criteria explored in
\citet{Busetto09}. \citet{ThorbergssonHooker} explores maximizing the
Fisher Information for a partially observed Markov decision process,
although this exacerbates the computational difficulties discussed
above.  Another direction for extension
  is to systems with vector parameters:our methods can be
readily extended to maximizing the trace of a Fisher Information matrix
($T$-optimal designs); other criteria such as the
determinant ($D$-optimality) that can be expressed as combinations of
time integrals can also be employed using the same
ideas.

We have dealt with parameter uncertainty before the experiment by
averaging the Fisher Information over the prior within our objective.
Alternative strategies such as maximizing the minimum Fisher Information
in a range of parameters can be implemented within the numerical control
strategy.  We have also not experimented with updating our prior as the
experiment progresses; this can be accomplished within particle
filtering, for example by treating parameters as additional state
variables -- an approach taken in \citet{Ionides06} -- although when
observations are highly informative about the system state, this can
result in particle filter collapse
\citep{Snyder08} (recent advances such as implicit filters
  \citep{ChorinMorzfeld12} may be helpful here).  \
Such an approach would also require either re-computing the
  optimal policy after each observation, or precomputing the control for
  a range of parameters and suitably combining them, but when
  the optimal control policy depends strongly on the system
parameters such updates may be expected to produce significant
improvements.

\paragraph{Acknowledgments.}
We thank Prof.~Cindy Greenwood for many helpful comments throughout the
initial stages of this work, and Prof.~Xueying Wang for help with the
Morris-Lecar model. We are also grateful to the anonymous
  referees for numerous suggestions and relevant references.  This work
was supported in part by the Statistical and Applied Mathematical
Sciences Institute (KL, BR), the National Science Foundation under
grants DMS-1418775 (KL), DMS-1053252 and DEB-1353039 (GH).


\bibliographystyle{chicago}
\bibliography{design}

\appendix


\section{Markov chain approximation}
\label{app:details}

This section provides technical details of our numerical methods for
obtaining the optimal control policy.  As noted above, in order to
execute the dynamic programming strategy on a computer and tabluate the
resulting control policy $F$, it is necessary to discretize the state
space.  Here we follow a discretization strategy due to Kushner
\citep{Kushner71}.  The starting point is the time-$t$ distribution
$\rho(t,\cdot)$ of the diffusion $\xbold_t$, which satisfies the forward
equation
\begin{equation}
  \partial_t\rho(t,\xbold) + \sum_i \partial_{x_i} \big(\fbold(\xbold, \theta,
  F(\xbold,t))\rho(t,\xbold) \big) = \frac12 \sum_{i,j}
  \partial_{x_i}\partial_{x_j}
  \big(\Sigma_{ij}(\xbold)~\rho(t,\xbold)\big)~,
  \label{eq:forward}
\end{equation}
where $F$ is the optimal control plan described above in the limit
$\dt\to0~.$ We discretize the optimization problem (\ref{optcontrol}) as
follows: (i) we cover the relevant portions of state space with a finite
grid of points with spacing $h$; (ii) we then discetize
Eq.~(\ref{eq:forward}) in space and time by a finite difference
approximation on this grid with a timestep $\dt^h$ depending on $h$; and
(iii) we interpret the coefficients of the finite difference
approximation as the transition probabilities of a finite-state Markov
chain (whose states are the discrete grid points chosen in (i)) to
approximate the diffusion process.
The dynamic programming algorithm can then be applied directly to this finite
state Markov chain in a straightforward manner.

As an example of what we mean by (iii), consider the simple diffusion
equation $\partial_tu = \partial_x^2u$.  The standard finite difference
discretization of this PDE is
\begin{align*}
  u(t+\dt,x)
  &= u(t,x) + \frac{\dt}{h^2}\big(u(t,x+h) + u(t,x-h) - 2u(t,x)\big) \\
  &= (1-2\mu)u(t,x) + \mu(u(t,x+h) + u(t,x-h))~,
\end{align*}
where $\mu=\dt/h^2~.$ Thus, if $\mu\in(0,\sfrac{1}{2})$, we can
interpret the equation above as describing a discrete state-space Markov
chain whose states are $\{0, \pm h, \pm2h, \cdots\}$ with transition
probability $P(X_{k+1}=x|X_k=x) = 1-\mu$ and $P(X_{k+1}=x\pm1|X_k=x) =
\mu$~.  Note that the condition $0<\mu<1$ imposes an upper bound on the
stepsize, specifically $\dt\leq h^2/2$~. Timestep limitations like
  this are a general feature of finite difference discretization
  schemes, and a poorly-designed finite difference scheme can lead to a
  great increase in computational cost.

The main technical problem is thus step (ii), since we need to ensure
that the coefficients of the finite difference scheme can be interpreted
as probabilities.  In our implementation, we employ a ``split operator''
finite difference discretization.  Roughly speaking, this amounts to
discretizing the drift and diffusion terms in the SDE~(\ref{eq:sde})
separately, then composing the resulting difference schemes.  This is a
standard technique often used for the numerical solution of partial
differential equations \citep[see, e.g.,][]{strikwerda}.  In our test
examples, the use of such a split operator scheme permits us to use
larger grid spacings and timesteps while maintaining numerical
stability, and can significantly speed up overall running speeds.


In detail: first, suppose for simplicity that the state space in
Eq.~(\ref{eq:sde}) is $\R$ (generalization to higher dimensions is
straightforward, but with more complex notation).  For each $h>0$, let
${\cal G}_h$ denote the regular grid $h\Z$ with spacing $h$.  For the
moment, assume ${\cal G}_h$ extends to all of $\R$; boundary conditions
are discussed below.  The grid ${\cal G}_h$ will be the state space of
our discrete-time approximating Markov chain $\tilde{\xbold}_i$.  To
choose time step sizes, we fix a constant $\mu>0$ and an integer $r>0$,
and set the timestep to be $\dt^h = \mu h^2.$ Then our approximation can
be described as follows:
\begin{quote}

  \subheading{Step 1.} Suppose at time $t=i\dt^h$, the chain has state
  $\tilde{\xbold}_i = sh.$ Then
  \begin{itemize}

  \item[-] with probability $|f(sh)|\cdot\dt^h/h$, jump to $(s+1)h$ if
    $f(sh)>0$ and to $(s-1)h$ if $f(sh)<0$; and

  \item[-] stay at $sh$ with probability $1-|f(sh)|\cdot\dt^h/h.$

  \end{itemize}

  \subheading{Step 2.}  Let $s'h$ denote the state of the chain after
  the previous step.  Then jump to $(s'\pm r)h$ with probability
  $\tfrac12\Sigma(s'h)\mu/r^2$, and stay at $s'h$ with probability
  $1-\Sigma(s'h)\mu/r^2$.

\end{quote}
This discretization scheme treats the drift and diffusion terms in
Eq.~(\ref{eq:sde}) separately, hence the name ``split operator.''  It is
straightforward to show that the composite finite difference
discretization provides a consistent discretization of
Eq.~(\ref{eq:forward}).  Step 1 interprets the drift term as a biased
jump, and Step 2 interprets the diffusion term as one step of a
symmetric random walk.

Both conceptually and in terms of programming, split operator schemes
like the one above are easy to work with; however they also impart an
additional variance on the motion of the approximating Markov chain.  In
the scheme above, this extra variance is on the order of $O(h\dt^h) =
O(h^3)$ per step; the relative effect of this extra variance (so-called
{\em numerical diffusion}) vanishes as $h\to0$.

Clearly, in order for the various expressions above to be valid
transition probabilities, we must have
$\mu/r^2<\inf_{x\in\R}\Sigma(x)^{-1}.$ We also need
$\sup_x|f(x)|\dt^h/h<1$.  Since $\dt^h/h = \mu h$, the second condition
can always be achieved by taking $h$ sufficiently small.  The first
condition, however, places a rather stringent constraint on the timestep
$\dt^h.$ The purpose of introducing the ``skip factor'' $r$ is
precisely to partially alleviate this constraint, at the cost of losing
some accuracy.  In practice, going from $r=1$ to $r=2$ can have a large
impact on the overall running time.  If $r=1$, our scheme closely
resembles the ``up-wind'' scheme of Kushner \citep{Kushner71}; a scheme
similar to our $r=2$ case is also described there.

In practice, rather than choosing $h$ and $\mu$, we usually first choose
$h$ and $\dt$, and take $r\propto\Sigma_{max}\sqrt{\mu}$ with a constant
of proportionality between 1 and 2, where $\Sigma_{max}$ is the maximum
of $\Sigma(x)$ over the domain of interest.  This guarantees that the
scheme converges (see below).  We usually use relatively coarse grids,
as (i) the structure of the optimal control policy is not too fine, and
(ii) we usually operate in the presence of observation and dynamical
noise, so there is not much point trying to pin down fine structures in
the control policy.

\subheading{Convergence.}  In general, a sufficient condition for the
Markov chain approximation method to be valid is that the approximating
chain $\tilde{\xbold}_i$ satisfy
\begin{align*}
  E_i(\tilde{\xbold}_{i+1} - \tilde{\xbold}_{i}) &=
  \fbold(\tilde{\xbold}_i, u_i)\dt + O(h^\alpha\dt)\\
  \var_i(\tilde{\xbold}_{i+1} - \tilde{\xbold}_{i}) &=
  \Sigma(\tilde{\xbold}_i)\dt + O(h^\alpha\dt)\\
  |\tilde{\xbold}_{i+1} - \tilde{\xbold}_{i}| &= O(h)
\end{align*}
\citep{Kushner71}. In the above, $E_i(\cdot)$ denotes conditional expectation given all
information up to step $i$, $\var_i(\cdot)$ denotes the corresponding
autocovariance matrix, and $\alpha>0$ is a constant.  Note the last line
should be interpreted to hold surely, i.e., it says the discrete
Markov chain makes jumps of $O(h)$ in size.  If the conditions above are
satisfied, then as $h\to0$, the optimal control policy for the
approximating chain will converge to an optimal control policy for the
diffusion~(\ref{eq:sde}).  It is easy to see that our scheme satisfies
the convergence criteria above with $\alpha=1.$

\subheading{Boundary conditions.}  In practice, ${\cal G}_h$
must be a finite grid.  We assume that it spans a subset of $\R^d$
sufficiently large that on the timescale of interest, trajectories
have very low probability of exiting.  We then impose that when the
approximating Markov chain attempts to exit the domain, it is forced to
stay at its current state.  This gives a simple way to obtain a Markov
chain on bounded grids.

\subheading{Higher dimensions.}  The generalization of the preceding
scheme to higher dimensions is straightforward: we simply treat each
dimension separately and independently.

\section{Numerical Methods for Particle Filters and Maximum Likelihood Estimation}
\label{app:filter}

In this appendix we provide details of our particle filter methods to estimate both the state value given current observations and for maximum likelihood estimation. We discretize the
diffusion process using the standard Euler-Maruyama approximation
(\ref{eq:euler}), and incorporate observations via Bayes's formula.
In all the examples considered in this paper, the observations are linear functions of the state vector plus Gaussian error. This allows us to avoid weighting and resampling in the particle filter. Instead, we incorporate the observations within the Euler-Maruyama scheme \eqref{eq:euler} by first applying the drift $\fbold(\Xbold_i,\theta,u_i)~\dt$ and then sampling the Gaussian disturbance $\sqrt{\dt}~\Sigma^{1/2}(\Xbold_i) \epsilonbold_i$ conditional on the observation at time $i+1$. When multiple steps are taken between observations, this update is applied for the final step.
%
In practical terms, our particle filters are
  simpler and more robust than bootstrap filters, and do not differ
significantly from other popular filtering schemes, e.g., the ensemble
Kalman filter.

One additional implementation detail: when observations are infrequent,
we simply hold the control value constant between observations.  In
principle, we can also use the past information to extrapolate the state
trajectory between observations, and use a time-dependent control.  But
for the examples considered in this paper, the simpler strategy appears
sufficient; it is also more realistically applicable in
  some experiments.

\paragraph{Maximum likelihood estimates}
Another advantage of using particle filters is that we can easily
evaluate the likelihood as a bi-product.  Let $I$ be the relevant
parameter interval (see Sect.~\ref{sect:Parameter Dependence and
  Priors}), and for simplicity assume a flat prior over $I$. We
discretize $I$ into a discrete grid $I_M$ of size $M$.  For each
putative parameter value in $I_M,$ we run a particle filter and compute
an estimate of the log-likelihood using the formula \citep{Ionides06}
\[
  \mbox{log likelihood}
   =
  \sum_{k=0}^{n}\log\big(p(\ybold_k|\ybold_1,\cdots,\ybold_{k-1})\big)
  \approx
  \sum_{k=0}^{n}\log\left(\frac1N\sum_{i=1}^Np\big(\ybold_k\big|\xbold^{(i)}_k\big)\right)
\]
where $\{\xbold^{(i)}_k:i=1,\cdots,N\}$ denotes the ensemble of
particles at step $k$.  If the computed log-likelihood
curve takes on a maximum in the interior of the interval $I,$ we do a
standard 3-point quadratic interpolation around the maximum and find the
maximizer; we use this as the MLE.  On the rare occasions when the
log-likelihood curve does not have a maximum inside the interval $I,$ we
use the corresponding endpoint as the estimate (but record the estimate
as being ``out of range'').

One further ``trick'' can be used to improve the performance of the
grid-based MLE: for each simulation experiment, corresponding
``particles'' for different parameter values are driven using the same
sequences of Gaussian random numbers.  That is, if the $i$th particle in
the ensemble for the $j$th parameter value is in state $\xbold^{i,j}_k$
at timestep $k$, then the particle positions $\xbold^{i,1}_{k+1},
\cdots, \xbold^{i,M}_{k+1}$ at the next timestep (where $M=$ number of
points in the parameter grid) are generated from the current positions
$\xbold^{i,1}_{k}, \cdots, \xbold^{i,M}_{k}$ using the {\em same}
Gaussian random numbers.  (To ensure correct sampling, the particles
$\xbold^{1,j}_t, \cdots, \xbold^{N,j}_t$ for fixed $j$ still receive
independent Gaussian random numbers.)  This {\em same-noise coupling} or
{\em method of same paths} reduces the variance of the resulting
estimates \citep{asmussen-glynn}, and produces smooth likelihood curves.

For comparison purposes, we also study the full observation
  problem in some of the examples, i.e., assume the full state vector is
  continuously and noiselessly observed.
  For such problems, the log-likelihood associated with particular
  sample paths are straightforward to calculate directly, and filtering
  is not needed.

\section{ML Model Parameters}
\label{app:ml}


Here we briefly summarize the details of the ML model used in this
paper.  The interested reader is referred to
\citep{ermentrout-rinzel,termantrout} for more details.

Recall the (deterministic) ML equations
\begin{displaymath}
  \begin{array}{rl}
    C_m \dot{v} &= I(t) - g_{\rm leak}\cdot(v-E_{\rm leak}) -
    g_K~w\cdot(v-E_K) - g_{\rm Ca}~m_\infty(v)\cdot(v-E_{\rm Ca})
    \\[2ex]
    \dot{w} &= \phi\cdot\big(w_\infty(v) - w\big)/\tau_w(v)
  \end{array}
\end{displaymath}
As explained in Sect.~\ref{sec:neuro}, the ML model tracks the membrane
voltage $v$ and a gating variable $w$.  The constant $C_m$ is the
membrane capacitance, $\phi$ a timescale parameter, and $I(t)$ an
injected current.  The reversal potentials $E_{\{K,Ca,leak\}}$ and
associated conductances $g_{\{K,Ca,leak\}}$ characterize the ion
channels and their dependence on the membrane voltage (see below).

Spike generation depends crucially on the presence of voltage-sensitive
ion channels that are permeable only to specific types of ions.  The ML
equations include just two ionic currents, here denoted calcium and
potassium.  The voltage response of ion channels is governed by the
$w$-equation and the auxiliary functions $w_\infty$, $\tau_w$, and
$m_\infty$, which have the form
\begin{displaymath}
  \begin{array}{rcl}
    m_\infty(v) &=& \frac12\big[1+\tanh\big((v-v_1)/v_2\big)\big], \\[1.5ex]
    \tau_w(v) &=& 1/\cosh\big((v-v_3)/(2v_4)\big),\\[1.5ex]
    w_\infty(v) &=& \frac12\big[1+\tanh\big((v-v_3)/v_4\big)\big].
  \end{array}
\end{displaymath}
The function $m_\infty(v)$ models the action of the relatively fast
calcium ion channels; $v_{\rm Ca}$ is the ``reversal'' (bias) potential
for the calcium current and $g_{\rm Ca}$ the corresponding conductance.
The gating variable $w$ and the functions $\tau_w(v)$ and $w_\infty(v)$
model the dynamics of slower-acting potassium channels, with its own
reversal potential $v_{\rm K}$ and conductance $g_{\rm K}$.  The
constants $v_{\rm leak}$ and $g_{\rm leak}$ characterize the ``leakage''
current that is present even when the neuron is in a ``quiescent''
state.  The forms of $m_\infty$, $\tau_w$, and $w_\infty$ (as well as
the values of the $v_i$) can be obtained by fitting data, or reduction
from more biophysically-faithful models of Hodgkin-Huxley type (see,
e.g., \citep{termantrout}).

The precise parameter values used in this paper are summarized in
Table~\ref{tab:params}.  These are obtained from \citep{termantrout}.
Throughout, the noise intensities are $\beta_v=1$, $\beta_w=0.1$.

\begin{table}
  \begin{center}
    \begin{tabular}{c|c}
      Parameter & Value \\[0.5ex]\hline
      $I_0$ & 95 \\[0.5ex]
      $C_m$ & 20.0 \\[0.5ex]
      $g_{\rm Ca}$ & 4.41498308$^{(*)}$ \\[0.5ex]
      $g_K$ & 8.0 \\[0.5ex]
      $g_{\rm leak}$ & 2.0 \\[0.5ex]
      $\phi$ & 0.04 \\[0.5ex]
    \end{tabular}
    \qquad\begin{tabular}{c|c}
      Parameter & Value \\\hline
      $v_K$ & -84.0 \\
      $v_{\rm leak}$ & -60.0 \\
      $v_{\rm Ca}$ & 120.0 \\
      $v_1$ & -1.2 \\
      $v_2$ & 18.0 \\
      $v_3$ & 2.0 \\
      $v_4$ & 30.0 \\
    \end{tabular}
  \end{center}
  \caption{Parameter values used for the Morris-Lecar example.  (*) The ``target''
    value of $g_{Ca}$ is randomly generated from a certain range.}
  \label{tab:params}
\end{table}


\section{Chemostat Growth Models} \label{sec:chemostat}

Our second example comes from experimental ecology.  In these
experiments, a glass tank or chemostat is inoculated with a population
of algae. The tank is bubbled to ensure that the contents are mixed, and
to prevent oxygen deprivation.  A nitrogen-rich medium is continuously
injected while the contents of the tank are evacuated at the same rate
as the injection.  Algae are assumed to consume nitrogen in proportion
to their population size and nitrogen concentration, until the
population size stabilizes due to saturation.  For ecological models a
set of stochastic differential equations can be proposed with drift term
\begin{align}
dN_t & = \left(\delta(t) \eta_I(t) - \frac{ \rho C_t N_t }{ \kappa + N_t} - \delta(t)N_t\right)dt \label{eqn:chemo} \\
dC_t & = \left(\frac{ \chi \rho C_t N_T }{ \kappa + N_t } - \delta(t) C\right)dt \nonumber
\end{align}
where the model has a mechanistic interpretation for an infinite
population given in terms of:
\begin{description}

\item[$N_t$] ($\mu$ mol/l) represents the nitrogen concentration in the
  chemostat

\item[$C_t$] ($10^9$ cells per liter) gives the relative algal density

\item[$\delta(t)$] (percentage per day) is the dilution rate; i.e., the
  rate at which medium is injected and the chemostat evacuated

\item[$\eta_I(t)$] ($\mu$ mol/l) is the nitrogen concentration in medium

\item[$\rho$] ($\mu$ mol/$10^9$ cells) is the rate of algal consumption
  of nitrogen

\item[$\kappa$] ($\mu$ mol/l) is a half-saturation constant indicating
  the value of $N$ at which $N/(\kappa + N)$ is half-way to its
  asymptote

\item[$\chi$] ($10^9$ cells/$\mu$ mol) is the algal conversion rate; how
  fast algae turn consumed nitrogen into new algae.

\end{description}
This system is made stochastic by multiplicative log-normal noise. This
is equivalent to a diffusion process on $N^*_t = \log(N_t)$ and $C^*_t =
\log(C_t)$ with additive noise:
\begin{align*}
dN^*_t & = \left(\delta(t) \eta_I(t) - \frac{ \rho e^{C^*_t+N^*_t} }{
  \kappa + e^{N^*_t}} - \delta(t)e^{N^*_t}\right)e^{-N^*_t}dt + \sigma_1
dW_1 \\
dC^*_t & = \left(\frac{ \chi \rho e^{C^*_t+N^*_t} }{ \kappa + e^{N^*_t}}
- \delta(t)e^{C^*_t} \right)e^{-C^*_t}dt + \sigma_2 dW_2.
\end{align*}
Here, the diffusion terms provide an approximation to stochastic
variation for large but finite populations and account for other sources
of extra-demographic variation.  While alternative parametrizations of
stochastic evolution can be employed, the diffusion approximation is
convenient for our purposes. A typical goal is to estimate the algal
conversion rate $\chi$ and the half-saturation constant $\kappa$.

This experiment involves several experimental parameters that can be
used as the dynamic input:
\begin{itemize}

\item The dilution rate of the chemostat $\delta(t)$

\item The concentration of nitrogen input $\eta_I(t)$

\item The times at which the samples are taken from the chemostat

\item The quantities that are observed.

\end{itemize}
In this paper we use $\delta(t)$ as the control parameter, which fits
well within the design methodology outlined above.  Realistic values for
the parameters, estimated from previous experiments, are
$(\eta_I,\rho,\chi,\kappa) = (160,270,0.0027,4.4)$. We chose $\sigma_1 =
\sigma_2 = 0.1$ based on visual agreement with past experiments; since
these act multiplicatively on the dynamics it is reasonable for them to
be of the same order of magnitude.  We hold $\eta_I(t)$ constant; within
the experimental apparatus, $\eta_I(t)$ can only be modified by changing
between discrete sources of medium.  Chemostats are typically inoculated
with a small number of algal cells and the models above give algal
density relative to a total population carrying capacity in the tens of
millions.  Under these circumstances, initial conditions can be given by
$N^*_0 = \log \eta_I$ (the input concentration) and $C^*_0 \propto \log
10^{-5}$ representing an initial population of a few hundred cells.  The
experimental apparatus places further constraints on the problem: the
dilution rate must be positive, and cannot exceed the maximal rate at
which medium can be pumped through the chemostat.

We focus on the estimation of $\kappa.$ In this model system, one can
measure both $C_t$ and $N_t$, though nitrogen measurements are both more
costly to make and prone to distortion by various factors.  Continuous
measurements are not practical for either quantity; one can at best make
measurements a few times a day.  Here, we focus mainly on the
complete-but-infrequenth observation regime, where we assume we can make
(noisy) measurements of both state variables.

\begin{figure}

  \begin{center}
    \includegraphics[bb=0in 0in 4in 3in,scale=0.8]{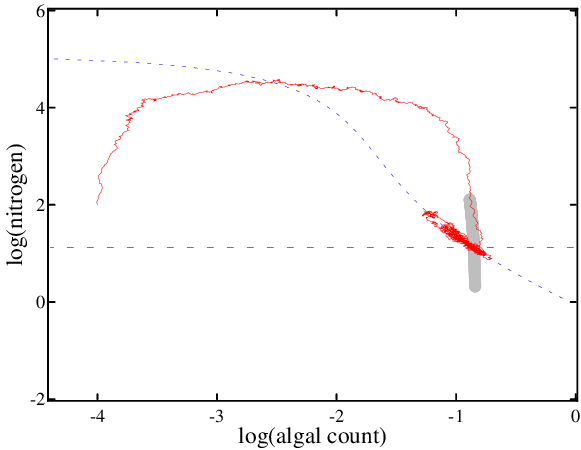}



  \end{center}

  \caption{State space plot of the chemostat model.  The solid curve is a
    single trajectory for the uncontrolled system.  The horizontal
    dashed lines are the nullclines of the noise-free system.  The thick
    gray line shows the movement of the fixed point as $\kappa$ varies
    from 2 to 12.  Here, $\delta=0.3$ and $\kappa=4.4~.$}
  \label{fig:chemo-phase}
\end{figure}

This system represents a most-marginal case for adaptive designs in that the system has a stable
fixed point for each dilution rate and most information is gained near the fixed point. Further, the value of $\kappa$
mostly affects the fixed point of $N_t$. We thus expect that adaptive controls will exhibit little improvement over
the best choice of a constant control policy and that estimates based solely on $C_t$ -- by far the easier quantity to measure -- will have poor statistical properties whatever policy is used. Figure \ref{fig:chemo-phase} shows an example trajectory of this model along with the fixed point as a function of $\kappa$.


\paragraph{Model properties and further parameter constraints}
We begin by examining the dynamic properties of the chemostat system.
In the absence of noise, i.e., when
$\sigma_1=\sigma_2=0~$, the phase space is fairly simple: it has a
single fixed point given by
\begin{equation}
  N^*_0 = \log\Big(\frac{\kappa\delta}{\chi\rho-\delta}\Big),~~~~
  C^*_0 = \log\Big(\frac{\delta(\eta_I-e^{N^*_0})(\kappa +
    e^{N^*_0})}{\rho e^{N^*_0}}\Big).
\end{equation}
It is easy to check that this fixed point is linearly stable, and
attracts all initial conditions in the relevant part of state space.
When $\sigma_1, \sigma_2>0,$ all trajectories eventually converge to a
neighborhood of the fixed point, and the stationary distribution is
concentrated around the fixed point for moderate $\sigma_1$, $\sigma_2$.
See Fig.~\ref{fig:chemo-phase}.  The fixed point exists only if the
dilution rate $\delta$ is not too high; when $\delta>\delta_c$ for some
critical $\delta_c,$ the fixed point $(C_0^*,N_0^*)$ moves to $-\infty.$
Physically, this means the dilution rate is so high that the algal
population crashes to 0.  For $\kappa\approx 4$ (the typical value we
assume in this example), the critical $\delta_c$ is roughly $0.7.$ To
avoid crashing populations, we keep the dilution rate below this number.

As mentioned above, $C^*_t$ is easier to measure experimentally because
it can potentially be done by an optical particle counter.
Unfortunately, in the parameter regime of interest, it is actually
fairly difficult to infer $\kappa$ from $C^*_t$ measurements alone.  To
see this, observe that for $\delta\in[0.1,0.6]$ and as $\kappa$ varies
over $[2,12],$ the fixed point moves essentially vertically; see
Fig.~\ref{fig:chemo-phase}.  Since trajectories are expected to stay
near the fixed point, it follows that changing $\kappa$ will have a
significant, observable impact on the $N^*_t$ dynamics but relatively
little on $C^*_t.$ Equivalently, because our estimation strategy is
based on the likelihoods of state variable observations, $N^*_t$
measurements (or combined $(C^*_t,N^*_t)$ measurements) will be much
more useful for estimating $\kappa$ than $C^*_t$ alone.  This is our
reason for focusing on the complete-observation case.

\begin{figure}

  \begin{center}
    \begin{small}
      \begin{tabular}{ccc}
        \includegraphics[height=3.1cm]{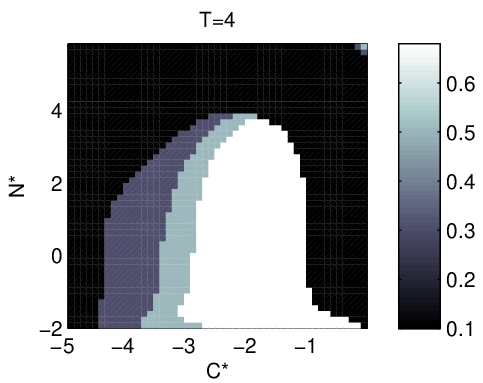} &
        \includegraphics[height=3.1cm]{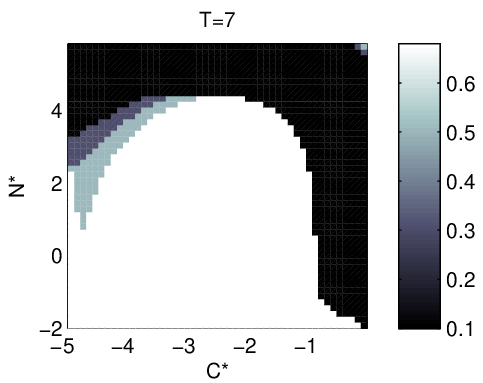} &
        \includegraphics[height=3.1cm]{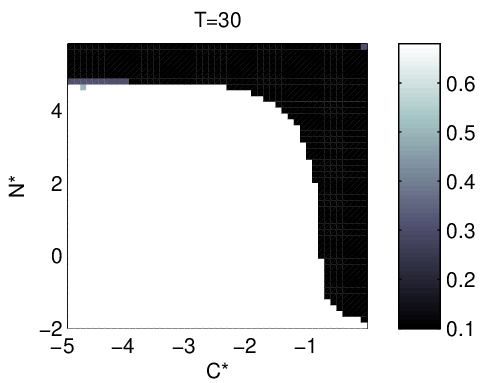} \\
        (a) $T=4$ & (b) $T=7$ & (c) $T=30$ \\[2ex]
      \end{tabular}
    \end{small}
  \end{center}

  \caption{Optimal control policies for the chemostat model.  In (a-c), we show the optimal
    control policies for different durations $T$. }
  \label{fig:chemo-controls}
\end{figure}

Finally, we note that the FI for $\kappa$ is given by
\[
 \left|\left| \frac{d}{d\theta}
\fbold(\xbold_t,\theta,u_t) \right|\right|^2_{\Sigma(\xbold_t)} = \frac{ \left(\frac{1}{\sigma^2_1}+\frac{\chi^2}{\sigma^2_2}\right) \rho^2 C_t^2
N_t^2}{\left( \kappa + N_t\right)^4}
\]
which is maximized when $N_t = \kappa$ but for $C_t \rightarrow
\infty~.$ As discussed in Sect.~\ref{sect:Parameter Dependence and
  Priors}, if we allowed unbounded control values, a penalization
would need to be imposed in order to avoid unrealistically large
diluation rates.  Since we are choosing dilution rates from a finite set
here, this is not necessary. Figure \ref{fig:chemo-controls} presents the control policy at
times $T$ = 4, 7 and 30.


\paragraph{Simulation results}
We now examine the effectiveness of the optimal control methodology
via simulations of the chemostat model.  We start at $(C^*_0,
N^*_0)=(-4,2)$, corresponding to typical experimental conditions.  With
this initial condition, it typically takes a trajectory about 7 days or
so (physical time) to reach equilibrium.  In our simulations, we carry
out experiments of duration $T$ for various values of $T$, ranging from
4 to 30 days. All simulation results are based on 256 realizations of
the (controlled) diffusion process.

\begin{figure}

  \begin{center}
%
%
%
%
%
%
%
    \begin{tabular}{cc}
    \includegraphics[height=3.7cm]{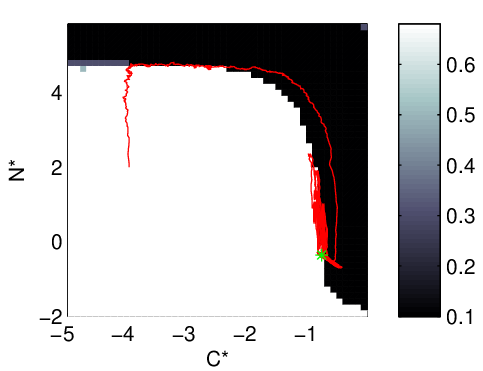} &
    \includegraphics*[height=3.7cm]{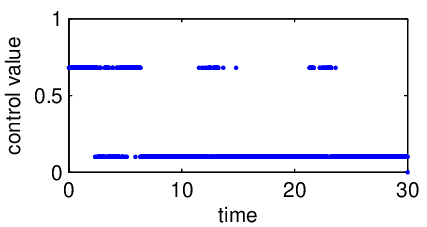} \\
    (a) A controlled trajectory & (b) The control $u_t$ \\
    \includegraphics[height=3.7cm]{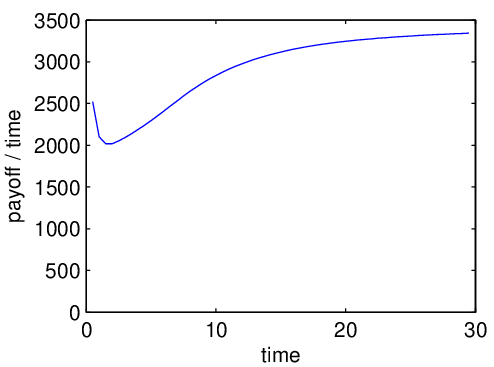} &
    \includegraphics[height=3.7cm]{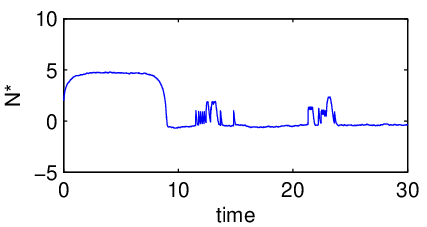} \\
    (c) Payoff / $T$ versus $T$ & (d) The controled $N^*_t$ \\
    \end{tabular}
  \end{center}

  \caption{(a) Controlled trajectories of the chemostat system \eqref{eqn:chemo} along with the control used (panel b) and the resulting $N^*_t$ (d).  Panel (c) shows
    $\overline{FI}(T) / T$, where $\overline{FI}(T)$ is the mean
    ``pay-off,'' i.e., Fisher information averaged over initial
    conditions and experimental realizations.}
  \label{fig:chemo-traj}
\end{figure}

%
%
The optimal control policies, for different values of $T$, are shown in
Fig.~\ref{fig:chemo-controls}.  The control values are drawn from the
set $\{0.1, 0.3, 0.5, 0.68\}$ (remaining below the dilution rate at
which all the algae is eventually removed from the system), and we
assume a flat prior for $\kappa$ over the interval $[3.5,5.5].$ As can
be seen, the computed control policy is nontrivial and depends on $T.$
However, we observed that after about $T\approx 10$, the control policy
stops changing.  The ``long-time,'' steady-state control policy uses
only the extreme values 0.1 and 0.68; on shorter timescales, the control
policy uses all available values.

Fig.~\ref{fig:chemo-traj}(c) shows the quantity
$\overline{FI}(T)/T$, where $\overline{FI}(T)$ is the mean Fisher
information (averaged over experimental realizations and initial
conditions) as a function of duration $T.$ As can be seen, by $T=30$,
the system has begun to approach the asymptotic regime, in which we
expect $\overline{FI}(T)\propto T.$ To illustrate the effects of the
optimal control, a controlled trajectory is shown in
Fig.~\ref{fig:chemo-traj}(a).  The corresponding control values are
shown in Fig.~\ref{fig:chemo-traj}(b), as is the corresponding values of
$N^*_t$ in Fig.~\ref{fig:chemo-traj}(d) (not much of interest happens to $C^*_t$).

\begin{table}
  \begin{center}
    \begin{tabular}{c|c|c|c|c|c|c|c}
      T & Control & N & In-range & Mean & Bias & Std.~Dev. & Std.~Dev.~Err \\\hline\hline
      4 & Dynamic & 256 & 54\% & 4.555 & 0.1551 & 0.7837 & 0.03464 \\
      4 & 0.1 & 256 & 62\% & 4.658 & 0.2577 & 0.7157 & 0.03163 \\
      4 & 0.3 & 256 & 26\% & 4.514 & 0.1137 & 0.9043 & 0.03996 \\
      4 & 0.68 & 256 & 16\% & 4.624 & 0.2243 & 0.938 & 0.04145 \\\hline
      7 & Dynamic & 256 & 100\% & 4.402 & 0.002351 & 0.0395 & 0.001746 \\
      7 & 0.1 & 256 & 100\% & 4.399 & -0.001245 & 0.04593 & 0.00203 \\
      7 & 0.3 & 256 & 35\% & 4.679 & 0.2788 & 0.8459 & 0.03738 \\
      7 & 0.68 & 256 & 18\% & 4.463 & 0.06275 & 0.9384 & 0.04147 \\\hline
      15 & Dynamic & 256 & 100\% & 4.402 & 0.001752 & 0.00955 & 4.221e-4 \\
      15 & 0.1 & 256 & 100\% & 4.403 & 0.002518 & 0.01284 & 5.674e-4 \\
      15 & 0.3 & 256 & 100\% & 4.399 & -8.251e-4 & 0.01969 & 8.704e-4 \\
      15 & 0.68 & 256 & 19\% & 4.446 & 0.04579 & 0.9314 & 0.04116 \\\hline
      30 & Dynamic & 256 & 100\% & 4.402 & 0.001827 & 0.004443 & 1.964e-4 \\
      30 & 0.1 & 256 & 100\% & 4.402 & 0.001629 & 0.005766 & 2.548e-4 \\
      30 & 0.3 & 256 & 100\% & 4.403 & 0.002547 & 0.01075 & 4.752e-4 \\
      30 & 0.68 & 256 & 21\% & 4.532 & 0.1324 & 0.9216 & 0.04073 \\
    \end{tabular}
  \end{center}
  \caption{Full observation results for the chemostat model.  We perform $N$ independent trials
    for each experimental set-up.  {\em in-range} = number of trials where
    estimator falls within uniform prior; {\em bias} = difference between
    the mean over tries and the true value; {\em std. dev.} = standard
    deviation of the estimator; and {\em std.~dev.~err.} = our estimate of
    the standard error of the standard deviation.}
  \label{tab:chemo-perfo}
\end{table}


To assess the quality of $\kappa$ estimates, we estimated the variance
of of the MLE, $\hat{\kappa}$ in simulations.
Table~\ref{tab:chemo-perfo} shows the results these.  As can be seen, on
shorter timescales, dynamic and constant controls do not differ much in
terms of performance.  But, on longer timescales (e.g., $T=30$), the
optimal control performs significantly better than most constant control
values, with the exception of $\delta = 0.1$: with $\delta=0.1$, the
constant-control system achieves close to the performance of the dynamic
control.  The reason for this is that given enough time, trajectories in
this system all converge to the stable fixed point.  At and near this
fixed point, the optimal policy sets $\delta = 0.1$, so once the
trajectory is close to the fixed point, the optimal policy is largely
indistinguishable from the constant policy.
%
%
In principle, the dynamic control can offer a potential gain in
information on shorter timescales (during the transient phase of the
dynamics).  But in this particular system, the dilution rate $\delta(t)$
has little effect on the dynamics during the transient, and controlled
and uncontrolled dynamics do not differ much.



%
%
%

\paragraph{Noisy and Partial Observation Simulation}
In Table~\ref{tab:chemo-fo}, we turn our attention to
complete-but-infrequent observation runs.  Here, we observe the system
twice a day (i.e., at intervals of $\Delta{t} = 0.5$).  Observation
noise is $(0.025, 0.025)$.  Naturally, all the standard deviations are
larger than the full observation case.  The overall trends observed for full
observations persist, and for the same reasons.

\begin{table}
  \begin{center}
    \begin{tabular}{c|c|c|c|c|c|c|c}
      T & Control & N & In-range & Mean & Bias & Std.~Dev. & Std.~Dev.~Err. \\\hline\hline
      4 & Dynamic & 256 & 41\% & 4.64 & 0.2398 & 0.8295 & 0.03666 \\
      4 & 0.1 & 256 & 49\% & 4.57 & 0.1696 & 0.7967 & 0.03521 \\
      4 & 0.3 & 256 & 22\% & 4.502 & 0.1024 & 0.9191 & 0.04062 \\
      4 & 0.68 & 256 & 18\% & 4.461 & 0.06094 & 0.9385 & 0.04148 \\\hline
      7 & Dynamic & 256 & 100\% & 4.415 & 0.01506 & 0.1332 & 0.005888 \\
      7 & 0.1 & 256 & 100\% & 4.398 & -0.002385 & 0.1416 & 0.006259 \\
      7 & 0.3 & 256 & 30\% & 4.619 & 0.2187 & 0.8772 & 0.03877 \\
      7 & 0.68 & 256 & 12\% & 4.463 & 0.06347 & 0.9587 & 0.04237 \\\hline
      15 & Dynamic & 256 & 100\% & 4.398 & -0.001966 & 0.06015 & 0.002658 \\
      15 & 0.1 & 256 & 100\% & 4.4 & -1.926e-4 & 0.05113 & 0.00226 \\
      15 & 0.3 & 256 & 100\% & 4.406 & 0.005561 & 0.08012 & 0.003541 \\
      15 & 0.68 & 256 & 20\% & 4.518 & 0.1175 & 0.9288 & 0.04105 \\\hline
      30 & Dynamic & 256 & 100\% & 4.4 & 2.777e-4 & 0.03385 & 0.001496 \\
      30 & 0.1 & 256 & 100\% & 4.399 & -6.979e-4 & 0.03149 & 0.001392 \\
      30 & 0.3 & 256 & 100\% & 4.406 & 0.005991 & 0.04461 & 0.001972 \\
      30 & 0.68 & 256 & 25\% & 4.446 & 0.04572 & 0.9107 & 0.04025 \\
    \end{tabular}
  \end{center}
  \caption{Complete but infrequent observation results for the chemostat model.  See
    Table~\ref{tab:chemo-perfo} for key.  }
  \label{tab:chemo-fo}

\end{table}
\begin{table}
  \begin{center}
    \begin{tabular}{c|c|c|c|c|c|c|c}
      T & Control & N & In-range & Mean & Bias & Std.~Dev. & Std.~Dev.~Err. \\\hline\hline
      4  & Dynamic & 256 & 79\% & 8.493 & 0.09294 & 2.902 & 0.1283 \\
      4  & 0.1 & 256 & 78\% & 8.51 & 0.1103 & 2.608 & 0.1153 \\
      4  & 0.3 & 256 & 58\% & 8.419 & 0.01897 & 3.481 & 0.1538 \\
      4  & 0.5 & 256 & 54\% & 7.952 & -0.4482 & 3.858 & 0.1705 \\
      4  & 0.7 & 256 & 58\% & 7.512 & -0.8882 & 3.798 & 0.1678 \\\hline
      7  & Dynamic & 256 & 72\% & 8.31 & -0.0896 & 3.175 & 0.1403 \\
      7  & 0.1 & 256 & 84\% & 8.383 & -0.01698 & 2.522 & 0.1114 \\
      7  & 0.3 & 256 & 70\% & 8.216 & -0.1837 & 3.116 & 0.1377 \\
      7  & 0.5 & 256 & 62\% & 7.922 & -0.4783 & 3.483 & 0.1539 \\
      7  & 0.7 & 256 & 55\% & 8.155 & -0.2448 & 3.707 & 0.1638 \\\hline
      15  & Dynamic & 256 & 73\% & 7.787 & -0.6127 & 3.288 & 0.1453 \\
      15  & 0.1 & 256 & 88\% & 8.168 & -0.2324 & 2.428 & 0.1073 \\
      15  & 0.3 & 256 & 78\% & 8.042 & -0.3575 & 2.999 & 0.1325 \\
      15  & 0.5 & 256 & 74\% & 8.144 & -0.2559 & 3.143 & 0.1389 \\
      15  & 0.7 & 256 & 65\% & 8.143 & -0.2572 & 3.494 & 0.1544 \\\hline
      30  & Dynamic & 256 & 84\% & 8.146 & -0.2544 & 2.678 & 0.1183 \\
      30  & 0.1 & 256 & 82\% & 8.277 & -0.1228 & 2.632 & 0.1163 \\
      30  & 0.3 & 256 & 77\% & 8.281 & -0.1193 & 3.006 & 0.1328 \\
      30  & 0.5 & 256 & 90\% & 7.933 & -0.4665 & 2.466 & 0.109 \\
      30  & 0.7 & 256 & 77\% & 8.154 & -0.2457 & 3.045 & 0.1346 \\
    \end{tabular}
  \end{center}
  \caption{Partial observation results.}
  \label{tab:chemo-po}

\end{table}

Finally, to confirm our intuition, we show the results of partial
observation runs in Table~\ref{tab:chemo-po}, where we observe only
$C^*_t.$ As expected, we cannot estimate $\kappa$ very well at all: many
of the estimates fall outside the prior range.  Nevertheless, if we
measure the estimator performance by the number of estimates that fall
inside the prior range, we see that the controlled system is comparable
to the best constant control.  In order to accommodate the larger
variance, we enlarge the prior range to $[2,12]$ in this set of
simulations, and also set $\kappa=8.4$.  This prior range is used both
to calculate the control policy and to obtain maximum likelihood
estimates as described in Sect.~\ref{sect:Filtering and estimation for
  partially observed systems}; since we cap the estimates to the
extremes of the prior range when the MLE falls outside the range, this
choice affects the estimator variance reported here.  We note that as
for any experimental design problem, in experimental situations the
prior must be determined beforehand and obtaining an MLE outside the
prior range is a good indicator that the prior was poorly chosen or the
data are uninformative about the parameter.

\end{document}